\documentclass[useAMS,usegraphicx,usenatbib]{mn2e}
\usepackage[T1]{fontenc}
\usepackage{aecompl}
\usepackage{amssymb}
\usepackage{breqn}

\newcommand{\msunyr}{M$_\odot$~yr$^{-1}$}
\newcommand{\msun}{M$_\odot$}
\newcommand{\rsun}{R$_\odot$}

\def\lesssim{\mathrel{\hbox{\rlap{\hbox{\lower3pt\hbox{$\sim$}}}\hbox{\raise2pt\hbox{$<$}}}}}
\def\gtrsim{\mathrel{\hbox{\rlap{\hbox{\lower3pt\hbox{$\sim$}}}\hbox{\raise2pt\hbox{$>$}}}}}
\def\gtreq{\mathrel{\hbox{\rlap{\hbox{\lower3pt\hbox{$-$}}}\hbox{\raise2pt\hbox{$>$}}}}}

\title[Hydrodynamic simulations of eccentric binaries]{Hydrodynamic Simulations of the Interaction between an AGB Star and a
Main Sequence Companion in Eccentric Orbits}
\author[Staff et al.]
{Jan E. Staff$^{1,2}$\thanks{E-mail: jan.staff@mq.edu.au}, Orsola De Marco$^{1,2}$, 
Daniel Macdonald$^{1,2,3}$, 
\newauthor Pablo Galaviz$^{1,2}$, Jean-Claude Passy$^4$, Roberto
Iaconi$^{1,2}$ and Mordecai-Mark Mac
Low$^{5,6}$\\
$^1$Department of Physics and Astronomy, Macquarie University NSW
2109, Australia\\
$^2$Astronomy, Astrophysics and Astrophotonics Research Centre, Macquarie University, Sydney, NSW 2109, Australia\\
$^3$Laboratoire Lagrange, UMR7293, Universit{\'e} de Nice Sophia-Antipolis,
CNRS, Observatoire de la C{\^o}te d'Azur, \\
06304 Nice Cedex 4, France\\
$^4$Argelander Institut f{\"u}r Astronomie, 53121 Bonn, Germany\\
$^5$Department of Astrophysics, American Museum of Natural History, New
York, NY, 10024, USA\\
$^6$Zentrum f{\"u}r Astronomie der Universit{\"a}t Heidelberg, Institut
f{\"u}r Theoretische Astrophysik, 69121 Heidelberg, Germany}
\begin{document}

\maketitle

\begin{abstract}

The Rotten Egg Nebula has at its core a binary composed of a Mira star and
an A-type companion at a separation $>$10~au.  It has been hypothesized to
have formed by strong binary interactions between the Mira and a companion
in an eccentric orbit during periastron passage $\sim$800 years ago.  We
have performed hydrodynamic simulations of an asymptotic giant branch star
interacting with companions with a range of masses in orbits with a range of
initial eccentricities and periastron separations.  For reasonable values of
the eccentricity, we find that Roche lobe overflow can take place only if
the periods are $\ll 100$~yr.  Moreover, mass transfer causes the system to
enter a common envelope phase within several orbits.  Since the central star
of the Rotten Egg nebula is an AGB star, we conclude that such a common
envelope phase must have lead to a merger, so the observed companion must
have been a tertiary companion of a binary that merged at the time of nebula
ejection.  Based on the mass and timescale of the simulated disc formed
around the companion before the common envelope phase, we analytically
estimate the properties of jets that could be launched.  Allowing for
super-Eddington accretion rates, we find that jets similar to those observed
are plausible, provided that the putative lost companion was relatively
massive.

\end{abstract}

\begin{keywords}
hydrodynamics --
methods: numerical -- 
binaries: close --
stars: evolution
\end{keywords}

\section{Introduction}

OH 231.8+4.2 is a binary consisting of an asymptotic giant branch (AGB) star
with a progenitor zero-age main sequence mass of 
$\sim3~{\rm M_\odot}$ \citep[found from cluster membership;][]{jura85} orbited 
by a
companion that is most likely a less massive A0V main sequence star (with a
mass of $1.5 - 2.5 {\rm M_\odot}$; \citealt{sanchezcontreras04}). 
The orbital separation is poorly constrained. 
\citet{sanchezcontreras04} assumed the binary separation to be 
$50~{\rm au}$ due to the presence of SiO, OH,   
and ${\rm H_2O}$ maser emission. \citet{schwarz95} argued that it
must exceed $10$~au, or else the observed maser would be tidally disrupted. 
Surrounding the binary is a thick torus of gas and dust, while  two highly 
collimated lobes extend perpendicular to the torus.  The kinematic age of 
the lobes indicates that they were formed $\sim800~{\rm yrs}$ ago
\citep{bujarrabal2002}. Both 
the torus and lobes contain approximately half a solar mass of dust and gas. 
The energy and linear momentum contained in the collimated outflows was 
measured to be $3\times10^{46}$~erg and $3\times 10^{49}$~g~cm~s$^{-1}$, 
respectively \citep{bujarrabal2002,alcolea2001}.
One lobe is longer than the other and appears to be breaking out of a bubble 
structure.  The nebula has been nicknamed the Rotten Egg and
Calabash nebula.

\citet{soker12} suggested that the Rotten Egg nebula, as well as other nebulae with similar characteristics of collimated, high momentum ejecta (e.g., IRAS22036+5306), were formed by the
 interaction of the two stars in the binary during periastron passage. 
 During the short-lived interaction time, mass would accrete from the AGB star to
 the main sequence companion and an accretion disc would form.  This disc 
 could launch jets that interact with the circumstellar
 environment leading to the inflation of lobes.  They also proposed that
 such an interaction could
repeat every periastron passage or every time a periastron passage coincides with an instability of the giant, such as
a thermal pulse in the case of an AGB star. \citet{kashi10} additionally suggested that the gravitational potential of the companion during periastron may be the actual trigger of such
an outburst on the primary.
Interactions such as these have been suggested to explain light outbursts such as
Intermediate Luminosity Optical Transients \citep[ILOTs;][also known as
Intermediate Luminosity Red Transients or Red Novae]{kashi10b} in isolation or
preceding more dramatic interactions such as common envelope (CE), leading to compact binaries or 
mergers.

It may be argued that giant stars should never be found in binaries with
eccentric orbits, because tidal circularisation time scales are relatively
rapid \citep[$\lesssim 1000~{\rm years}$;][]{soker00}.  However, eccentric
binaries with relatively small orbital separations do exist.  A high 10 per
cent of the sequence E stars in the Large Magellanic Cloud (LMC) are such
binaries \citep{nicholls12}.  These stars are ellipsoidal variables where a
red giant branch (RGB) or an AGB star is in a close orbit with a main
sequence companion and the eccentricity has been observed to be as high as $\sim$0.4 \citep{nicholls12}. 
\citet{soker00} suggested that enhanced mass loss rate during periastron
passages could increase the eccentricity, explaining why such eccentric
systems exist.

Here we carry out a first study of the viability of the proposal that
binaries in eccentric orbits could interact strongly enough to promote the
repeated ejection of relatively massive collimated structures.  Accretion of
mass onto the companion from the {\it wind} of an AGB star has been studied
and deemed unlikely to drive the massive jets observed in nebulae like the
Rotten Egg Nebula \citep{HuarteEspinosa2013}.  We therefore carry out
hydrodynamic simulations of the accretion of mass onto a companion during
Roche lobe overflow (RLOF). Since we need a RLOF event to take place at
periastron, we exclude any orbit with a period $\gg100~{\rm yrs}$, since
otherwise the eccentricity would have to approach unity to allow a
periastron separation sufficiently short for a RLOF to occur.
 
Hydrodynamical simulations of interacting binaries are challenging,
particularly for eccentric orbits. A few groups have
simulated similar setups using a variety of techniques. \citet{church09} used an
``oil-on-water'' smoothed particle hydrodynamics (SPH) technique to simulate binaries consisting of a white
dwarf and a main sequence star with a range of eccentricities to measure the
semi-major axis required to start mass transfer. They found that for higher
eccentricity, the periastron separation must be smaller for mass transfer to commence. 
\citet{lajoie11} presented results
of SPH simulations of two main sequence stars in eccentric orbits. They
found episodes of accretion, with the rate peaking slightly after periastron passage. A
small amount of mass ($\sim5\%$) is lost from  the system in the
process. 
CE simulations between giants and their companions tend to start with the 
companion already at the surface of the giant, thereby bypassing the entire 
pre-contact and Roche lobe contact phase \citep[e.g.,][]{ricker12,passy12} 

In this paper we carry out a limited set of hydrodynamic simulations of
AGB stars with a companion in eccentric orbits and periods between $\sim10$
and $\sim100$ years.  We explore the
effect of varying the periastron separation, initial eccentricity and
companion mass. Since we consider the possibility that the A0V main sequence companion
was responsible for the interaction, we use a companion mass of $1.7~{\rm
M_\odot}$ in most cases. However, we also consider the possibility of a higher
and lower companion mass, in case the A0V star was not responsible for
the interaction but 
instead a second companion interacted and merged with the AGB star to form the 
present day nebula. We concentrate on whether a disc is formed around
the companion and whether this disc could drive the mass
ejections deduced for the Rotten Egg Nebula.  These simulations must be
considered approximate because they neglect several potentially important 
effects. However, they include sufficient complexity to enable
us to think critically about these proposed models and suggest improvements
and tests.

In particular, we will address the following questions:
(i) How much mass is unbound from the system? 
(ii) What periastron separations will cause significant mass transfer onto
the companion?
(iii) If mass is transferred, will the interaction lead to an accretion disc 
around the companion
that can launch a jet able to explain the lobes in the
Rotten Egg Nebula?
(iv) Will the interaction lead to a merger, and if not can the interaction 
result in a wide orbit binary as is observed in
the Rotten Egg Nebula?

In Section~\ref{sec:method} we describe the method and
initial setup for the simulations.  Then in Sec.  3 we
present the results: spinning up of the AGB star
(Sec.~3.1), mass and longevity of the accretion disc (Sec.~3.2), the
effect of a less and a more massive companion (Sec.~3.3),
evolution of the orbital separation (Sec.~3.4), mass lost from the
system (Sec.~3.5), and finally, numerical considerations (Sec.~3.6).  
In Sec.  4 we
discuss the type of magnetic fields and jets that could arise from such
discs. In Sec.  5 we discuss and conclude,  while in Sec. 6 we 
summarize our results.

\section{Method and initial setup}
\label{sec:method}

Using the 1-dimensional stellar evolution code \textsc{mesa} \citep[Modules for
Experiments in Stellar Astrophysics;][]{paxton11,paxton13}, we
calculated the structure of an AGB star with a mass of $3.05~{\rm M_\odot}$
(zero-age main sequence mass of $3.5~{\rm M_\odot}$, similar to that
deduced for OH 231.8+4.2 by \citealt{jura85}) and
solar metallicity. This object has an age of $\sim3\times10^8~{\rm
yrs}$, similar to the age of NGC 2437 which OH 231.8+4.2 belongs to
\citep{mermilliod81}. At this point in the stellar evolution, this AGB star
has a radius $R_{\rm AGB}\sim 2.2~{\rm au}$. This star is then imported into a modified version of
the 3-dimensional hydrodynamics code \textsc{enzo} \citep{oshea04,passy12,bryan14}.  This is
a grid based hydrodynamics code, using Cartesian coordinates. Using the Zeus module in \textsc{enzo}, we evolve the hydrodynamics
equations as in \citep{passy12}. In particular we evolve the internal energy
rather than the total energy, and we calculate the gravitational potential
using the Poisson equation.

We use a uniform grid with a resolution of $256^3$ and a domain size of
$4.4\times10^{14}~{\rm cm}$ in all three directions in most cases.  In one case we use
$512^3$ cells (simulation \#4hr in Table~1).  In two cases we simulate a
larger volume of $5.3\times10^{14}~{\rm cm}$ on a side and in order to keep
the same grid-cell size we use a grid with $306^3$ cells (simulations \#2 and
\#8 in
Table~\ref{initialsetups}).  The core of the AGB star is not resolved and is simulated as a
point mass, as is the companion star \citep[similarly to][]{passy12}.  The
AGB core point mass is $0.7~{\rm M_\odot}$, the same as the AGB star's
carbon-oxygen core mass.  In most of the simulations, the main sequence 
companion has a mass of
$1.7~{\rm M_\odot}$, consistent with the mass of an A0V main sequence star.
However, as it is possible that a third object was responsible for the
interaction and merged with the AGB star, we have also run a
simulation with a $0.6~{\rm M_\odot}$ companion and one with a $3~{\rm
M_\odot}$ companion, both with eccentricities of $0.7$, to investigate how
the companion's mass affects the results.

While \textsc{mesa} has a very high resolution in one dimension and takes
the microphysics into account, we use the 3-dimensional \textsc{enzo} code
with a much coarser resolution, and with an ideal gas equation of state with
$\gamma=5/3$.  As a consequence of the changed resolution, the use of a
particle as the stellar core, and the different equation of state, a stable
stellar model in \textsc{mesa} may not be stable in \textsc{enzo}.  After
mapping the stellar model from \textsc{mesa} into \textsc{enzo}, we fill the
surroundings with a low density (four orders of magnitude less than the
lowest density in the star), hot gas, to match the stellar pressure at the
surface.  While damping the velocity field by a factor of 3 for each cycle,
as done by \citet{passy12}, the star is evolved without a companion for
several dynamical timescales so that a stable object is formed.  Before
inserting the companion, we evolve the star for several dynamical times
without damping the velocities, to verify its stability.  We also test the
stability of the AGB structure in the \textsc{enzo} domain by evolving the
simulation with a single AGB star that is moving across the grid with a
velocity typical of the orbital velocity of the primary in the binary
system.

\begin{figure}
\includegraphics[width=0.45\textwidth]{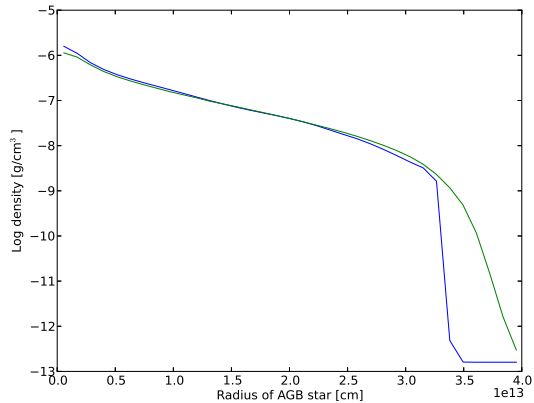}
\caption{ A comparison of the \textsc{mesa} AGB star density profile after mapping it into the \textsc{enzo} computational domain (blue curve) and the same profile 
after four years of relaxation when the structure stabilizes (green curve) 
for a simulation where the AGB star is single and moving across the grid, with a velocity similar to what it
would have initially if there was a companion present.}
\label{densityprofile}
\end{figure}

The \textsc{mesa} model outermost grid point is at the photosphere and has a
density of $\sim$$10^{-9}~{\rm g~cm^{-3}}$.  Hence, there is a sharp density
jump from the photospheric density to the ambient background density four
orders of magnitude lower.  Over the stabilisation time, {despite the
pressure support from the surrounding ambient medium,} the stellar model
diffuses out due to this sharp density jump (see Fig.~\ref{densityprofile}). 
After stabilisation, at a density of $10^{-10}~{\rm g~cm^{-3}}$, the star is
$\sim10$ per cent larger than initially. Being at a larger radius, this 
material is less tightly bound
and could more easily be dragged off during an interaction.
Since very little mass resides in this puffed-up, low density material, 
we do not, however, expect it to affect our simulations significantly. The 
low density
ambient medium causes a negligible drag on the companion as it moves through 
it.

The ambient medium has a high pressure, meant to equalize the outer part of
the star initially.  During the early part of the simulation, the ambient
medium can therefore prevent low pressure gas from expanding as much as it
would naturally.

We insert the point mass companion into the computational domain and both
stars (the giant with its core and the companion) are given velocities
appropriate for the desired orbit.  We do not stabilize the AGB star within
the potential of the companion. This is only possible 
in the co-rotating frame, which is hard to implement for non-circular orbits. 
We determined that if the companion is initially placed at a separation
$\gtrsim6~{\rm R_{AGB}}$, it does not significantly affect the structure of
the giant star. If the companion is inserted closer, the giant star is less
distorted than if it had been allowed to adjust to the altered potential. As a
consequence, the interaction during periastron passage in
eccentric simulations like these will be
weaker. This is primarily a problem in our less eccentric simulations
(i.e. simulation \#4, with an eccentricity of 0.33). This effect may
influence the details of the simulations, but is unlikely to affect the main
results, namely that the system soon enters a CE, and that little mass is
unbound.

The gravitational potential created by a point mass is smoothed to prevent a
singularity, as done by \citet{sandquist98}: 
\begin{equation}
\phi(r)={-{GM}\over{\sqrt{r^2+\epsilon^2 \delta^2 \exp{(-r^2/(\epsilon\delta)^2)}}}};
\end{equation} 
where $G$ is the gravitational constant, $M$ is the mass of the point particle, 
$r$ is the distance from it, $\delta$ is the size of a cell and $\epsilon$, the smoothing length,  has a value of 1.5 or 
$3$  (see
Table~\ref{initialsetups}). A smoothing length of 1.5 cells results in better resolution, but the potential gradients
can be too steep for satisfactory energy conservation (discussed further in Section~\ref{resultssection}).

\subsection{The Roche lobe radius}
\label{ssec:roche}

The aim of our simulation is to determine the type of interaction happening
at or around the time of RLOF for an eccentric binary.  Guided by the
kinematic parameters of the Rotten Egg Nebula, which seems to have suffered
an outburst $\sim$800 years ago, we attempted to model a relatively long
orbital period, but with a relatively short periastron distance.  This can
only be achieved with relatively high eccentricities. \citet{nicholls12}
found many high eccentricity,
close binary stars with a giant primary, but not with
an eccentricity value as high as we have selected for some of our
simulations ($e\simeq$0.7-0.9).  We discuss the implications of our selections in
Section~\ref{ssec:accretion}.

Using the expression for the effective Roche lobe radius from 
\citet{eggleton83}:
\begin{equation}
r_L=\frac{0.49q^{2/3}}{0.6q^{2/3}+\ln{[1+q^{1/3}]}},
\label{eggleton}
\end{equation}
with $q=M_{\rm d}/M_{\rm a}$ being the mass ratio of the two stars, we 
find that an $M_{\rm d}=3.05~{\rm M_\odot}$ donor star with an 
$M_{\rm a}=1.7~{\rm M_\odot}$ accreting companion ($q=1.79$) has an effective
Roche lobe radius of $\approx 0.43$ times the separation of the two stars.
Hence, for the $3.05~{\rm M_\odot}$ star to exactly fill its Roche lobe, the
separation of the two stars must be $2.3$ times the radius of the 
$3.05~{\rm M_\odot}$ star of $2.2~{\rm au}$.
This expression, however, assumes that the stars are rotating synchronously
and in circular orbits.
In our case, the stars are not rotating initially, and are not in circular
orbits. \citet{sepinsky07} found
that binary components rotating slower than the orbital velocity at periastron
(as is
likely to be our case at the first periastron passage) would have a Roche
lobe radius up to $10$ per cent larger than that given by the expression in Eq.~\ref{eggleton}. This can therefore reduce the mass transfer 
slightly. On the other hand, as we described above, the low density outer layers of the star do diffuse out slightly (by about $10$ per cent referring to a density contour of $10^{-10}~{\rm g~cm^{-3}}$; Fig.~\ref{densityprofile}), making the star bigger and hence more likely to overflow its Roche lobe.
These effects are opposite and may balance each other out to an extent.

\section{Results}
\label{resultssection}

\begin{table*}
\begin{minipage}{\textwidth}
\caption{Overview of the initial setup and results for the simulations 
discussed in this paper. In all cases, the mass of the AGB star is 
$3.05~{\rm M_\odot}$, its radius is ${\rm R_{AGB}} =2.2~{\rm au}$.
The simulation domain is a cube with sides
$4.4\times10^{14}$ cm long, except for simulations \#2 and \#8 (see the text), run 
at $256^3$ resolution (except for simulations \#2, \#8 and \#4hr). We refer to the
labels given in this table when discussing the simulations in the text. The
unbound material refers to the total unbound mass. The final separation and 
the accumulated unbound material are not available for simulations \#3 and 
\#6, since in these cases the
orbit following the first periastron passage was sufficiently large that the 
companion left the grid. For these simulations the period and eccentricity
have been estimated based on the velocities after periastron passage.}
\begin{tabular}{ccccccccccccc}
\hline
Label & Comp. & Initial & Initial & Initial & Initial & Smo- & Period & Eccen-
& Disc mass & Final & Unbound & Unbound\\
& mass & separ- & orbital & peri. & eccen- & othing & after first & tricity
& after first & sep. & material\footnote{Including thermal energy in
calculating the unbound mass} &
material\footnote{Not including thermal energy when calculating the unbound
mass}\\
& & ation &  period & sep. & tricity & length & peri.pass. & after & peri.pass.
& (Fig.~\ref{35sep}) \\ 
\# & (${\rm M_\odot}$) & (au) & (yr) & $({\rm R_{AGB}})$ & & (cells) & (yr) &
peri.pass. & (${\rm M_\odot}$) &
(au) &
(${\rm M_\odot}$) & (${\rm M_\odot}$) \\
\hline
1 & 1.7 & 22 & 25.8 & 2.0 & 0.70 & 3.0 & 14 & 0.60 & 0.04 & 0.4 & 0.28 & 0.13\\
1ss\footnote{$1.5$ cells smoothing length} & 1.7
& 22 & 25.8 & 2.0 & 0.70 & 1.5 & 12 & 0.62 & 0.02 & 0.4 & 0.31 & 0.14 \\
2\footnote{$306^3$ resolution; bigger box, same cell size} & 1.7 & 13.2 & 36.0 & 2.5 & 0.70 & 1.5 & 27 & 0.65 & 0.0004 & 0.6 & 0.35 & 0.12 \\
& & & & & & & 22 & 0.58 & $0.02$\footnote{After second periastron passage} & & \\
3 & 1.7 & 13.2 & 49.9 & 3.0 & 0.70 & 1.5 & 36 & 0.64 & 0.0 & - & - & -\\ 
4 & 1.7 & 8.4 & 7.7 & 2.0 & 0.33 & 1.5 & 5 & 0.25 & 0.05 & 0.4 & 0.42 & 0.15\\
4hr\footnote{$512^3$ resolution} & 1.7 & 8.4 & 7.7 & 2.0 & 0.33 &
3.0 & 6 & 0.29 & 0.03 & 0.2 & 0.16 & 0.12 \\
5 & 1.7 & 12.1 & 12.0 & 2.0 & 0.50 & 1.5 & 9 & 0.40 & 0.02 & 0.4 & 0.42 & 0.16 \\
6 & 1.7 & 13.2 & 134 & 2.0 & 0.90 & 1.5 & 55 & 0.80 & 0.002 & - & - & -\\
7 & 0.6 & 13.2 & 19.2 & 1.5 & 0.7 & 3.0 & 12 & 0.48 & 0.001 & 0.3 & 0.14 & 0.08\\
8 & 3.0 & 22 & 32.0 & 2.5 & 0.7 & 1.5 & 16 & 0.65 & 0.06 & 1.0 & 0.22 & 0.12 \\
\hline
\end{tabular}
\label{initialsetups}
\end{minipage}
\end{table*}

\begin{figure*}
\includegraphics[width=0.95\textwidth]{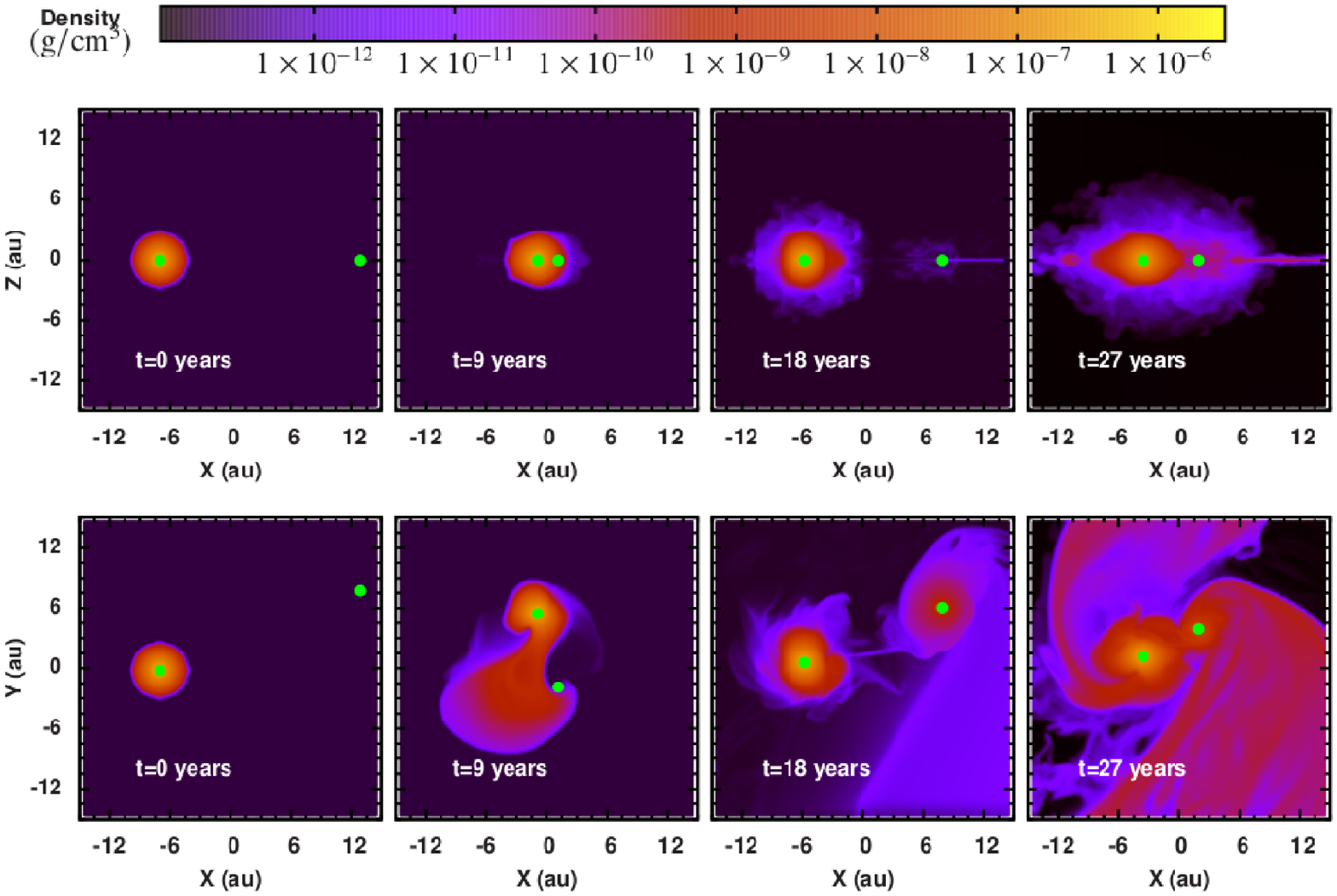}
\vspace{0.5cm}
\caption{Snapshots showing density slices perpendicular to 
the equatorial plane (top panel) and along the equatorial plane,
(bottom panels) at four different times in simulation \#1
(Table~\ref{initialsetups}). 
The green dots are the
particles positions projected onto the plane.}
\label{2rsfig}
\end{figure*}

We have carried out a number of simulations, whose initial configurations
are summarized in Table~\ref{initialsetups}.  We choose simulation \#1 (a
$3.05~{\rm M_\odot}$ AGB star with a $1.7~{\rm M_\odot}$ companion, initial
eccentricity of 0.7, a periastron separation of $2~{\rm R_{AGB}}$ and an
orbital period of $\sim$26 years) as an illustrative case and describe it
below.  In Sections \ref{ssec:rotation} to \ref{ssec:energy} we give a
quantitative account of the results and compare different simulations.

In Figure~\ref{2rsfig} we present snapshots taken at different times showing
density slices through the equatorial and perpendicular planes.  During the
first periastron passage, that takes place 8 years after the start of the
simulation, the AGB star overflows its Roche lobe and $0.04~{\rm M_\odot}$ is
lost to a disc around the companion, while about $0.02~{\rm M_\odot}$ is
lost from the simulation box (this mass is not necessarily unbound).  In
addition, a low density stream of gas forms between the AGB star and the
companion (see panel after $t=18~{\rm years}$ in Fig.~\ref{2rsfig}). 
The stream and the plume of gas trailing the companion that can be
seen in the figure after 18
years are confined to the midplane of the simulation 
as they are in pressure equilibrium with the hot ambient medium. In a more
realistic setup without the hot ambient medium, the plume would likely 
have been more puffed up.
The
companion's orbit is altered by the interaction and the companion with its
disc return to periastron after only about 14 years.  The AGB star is at this
point distorted (see Fig.~\ref{2rsfig}), and its 
diameter, out to a density of $10^{-9}~{\rm g~cm^{-3}}$, is now between 10 and
30 per cent larger than it was initially, with much lower density material
extending beyond that.

During the second periastron passage, more mass is dragged off the AGB star, 
some of which
flows out of the simulation box (about $0.05~{\rm M_\odot}$ is lost from the
simulation box during the next orbit as a consequence of this interaction),
while $0.05~{\rm M_\odot}$ are captured by the companion, so that the disc 
mass grows to $0.09~{\rm M_\odot}$. The
companion's orbit is altered further and the companion remains 
engulfed in the AGB star's envelope (see the panel at $t=27$ years in Fig.~\ref{2rsfig}). 

The CE proceeds in a manner qualitatively similar to previous
simulations \citep{sandquist98,passy12}.  We stop the simulation when
the orbital in-spiral slows down, as described by \citet{passy12}.  We call
mass that has a positive total energy unbound, and we calculate the total
energy by both including or excluding the thermal energy i.e., gas is
unbound if $E_{\rm k} + E_{\rm pot} + E_{\rm th} > 0$ or $E_{\rm k} + E_{\rm
pot} > 0$; note it has been argued that the enthalpy and not the thermal
energy should be used \citep{ivanova11}, but this detail has no
bearing on our conclusion.  Including the thermal energy likely overestimates the amount of
unbound mass.  However, it is possible that part of the thermal energy will
be converted into kinetic energy of the gas.  Hence ignoring the thermal
energy in the computation of the unbound mass may underestimate the amount
of unbound material.  Therefore we report the unbound mass calculated with
both methods in Table~\ref{initialsetups}.  Finally, unbinding of mass 
generally occurs
when gas interacts with the particles.  As unbound gas travels outwards it may
interact with bound material farther out, causing it to lose sufficient
energy that it becomes bound, making it possible for the total amount of
unbound gas on the grid to drop without mass being lost from the grid.

The high resolution simulation (\#4hr) ran over 8 days on 256 cores on the
National Computational Infrastructure's
Raijin\footnote{nci.org.au/nci-systems/national-facility/peak-system/raijin/} 
machine. Simulation \#2 was 
the longest simulation of the lower
resolution simulation set and took approximately one week on 128 cores. In
addition the stellar stabilization (section 2) took approximately two days.

\subsection{The AGB star's rotation}
\label{ssec:rotation}

The interaction with the companion transfers orbital angular momentum to the
AGB star.  All CE interactions must take place between a
companion and a spun-up star because a tidal interaction must have taken
place on a shorter timescale than the radial evolution of the binary,
meaning that orbital angular momentum is transferred to the primary by tidal dissipation as the
companion is brought into contact.  However, it is
non-trivial to simulate an interaction with a rotating giant because setting
up a stable, rotating star in the 3D computational domain requires, once
again, the co-rotating frame. In the co-rotating frame, there are no orbital
velocities. Any velocity can therefore be dampened as they arise from
instabilities. 

 In a rotating giant the velocity contrast between the companion and the
envelope gas is smaller, which will result in a lower gravitational drag
force experienced by the companion.  This will likely change the nature of
the interaction.  \citet{sandquist98} reported that setting their 3-\msun\
AGB primary in synchronous rotation with the $0.4~{\rm M_\odot}$ companion
at a distance of $2\times10^{13}~{\rm cm}$ (equivalent to $1.44~{\rm
R_{star}}$) did not alter the outcome of the CE interaction compared to a 
situation with no rotation.

We investigate the velocity field of the AGB star, in the stellar core's frame
of reference, after 20 years of simulation time, 12 years after the first
periastron passage, corresponding to 6.5 AGB stars dynamical times.  In order
to quantify the rotational velocity, we find the velocity at different
constant radii and constant densities in the equatorial plane in the
envelope.  The results are given in Table~\ref{velocitytable}.  The highest
velocities are found at low densities and large radii, as expected since
those layers suffered the strongest interaction.  How the velocity profile
will redistribute depends greatly on the stellar structure as well as its
magnetic fields \citep{Cantiello2014} and we cannot hope to reproduce it
here.  However, the amount of angular momentum received by the star from the
spin-orbit interaction is likely reasonably accurate at the angular momentum
conservation level observed.

\begin{table}
\caption{The velocity modulus in the equatorial plane of 
the AGB star's envelope, in the frame of reference of the core of the AGB star,  after
20 years in simulation \#1, at different radii.}
\begin{tabular}{cccc}
\hline
Radius & Density & Velocity & Mean velocity \\
(au) & $(10^{-10}{\rm g~cm^{-3}})$ & $({\rm km~s^{-1}})$ & $({\rm km~s^{-1}})$ \\ 
\hline
0.9 -- 1.1 & $500$ -- $10\,000$ & $0.4 - 3$ & $2$\\ 
1.7 -- 2.0 & $80$ -- $100$ & 2 -- 8 & $5$ \\
2.1 -- 2.3 & $2$ -- $80$ & 5 -- 15 & 8 \\
2.6 -- 3.0 & $8$ -- $10$ & 5 -- 20  & 12 \\
\hline
\end{tabular}
\label{velocitytable}
\end{table}

\begin{figure}
\includegraphics[width=0.49\textwidth]{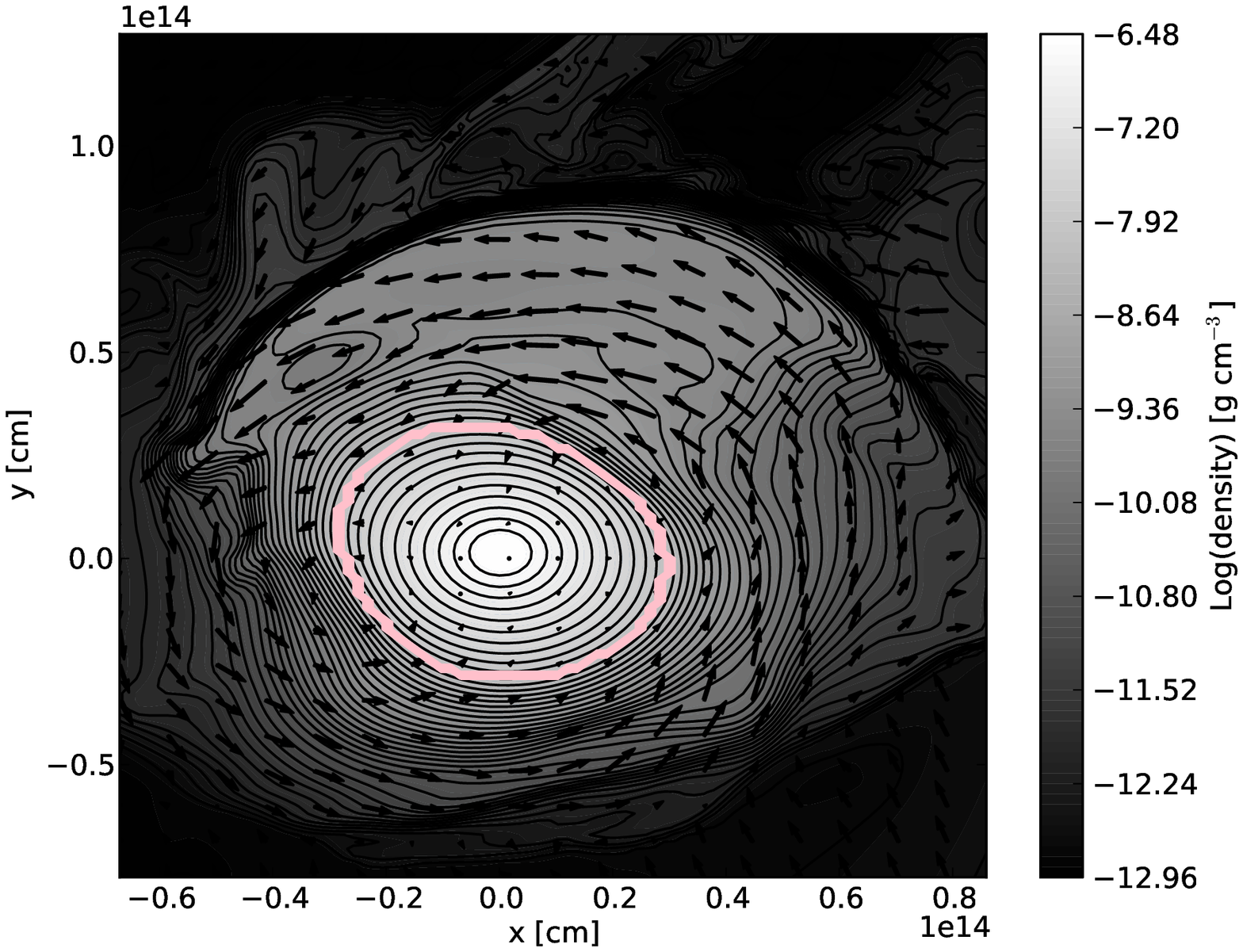}
\includegraphics[width=0.49\textwidth]{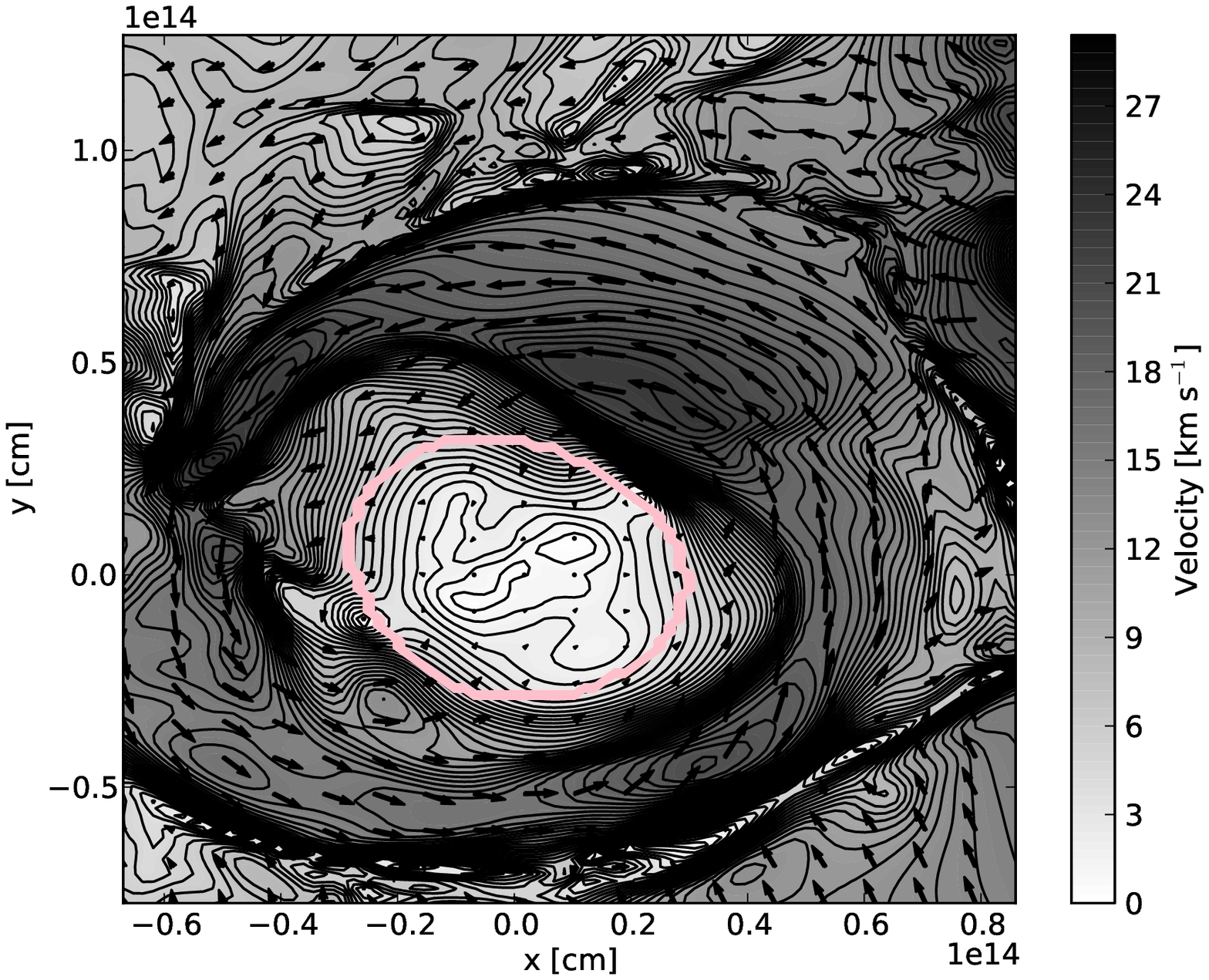}
\caption{The logarithm of density ({\it top}), and the absolute value of
the velocity vector field in the frame of reference of the AGB star's core 
({\it bottom}), in 
the orbital plane, 20 years into the
simulation (about 12 years after the first periastron passage), over-plotted 
with the velocity vectors on that plane, drawn every 5 zones in the grid. 
The pink curve indicates densities between $8\times10^{-9}~{\rm g~cm^{-3}}$ 
and $10^{-8}~{\rm g~cm^{-3}}$. The core of the AGB star is at (0,0) in the
figure.}
\label{velfieldfig}
\end{figure}

\begin{figure}
\includegraphics[width=0.49\textwidth]{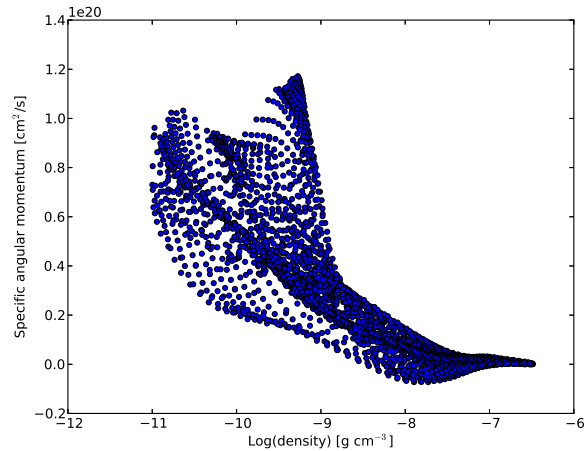}
\caption{The specific angular momentum of the AGB star in the orbital plane
as a function of the
logarithm of the gas density, for simulation \#1 after 20 years.
The slightly negative specific angular momenta found around
$10^{-8}~{\rm g~cm^{-3}}$ show counter
rotation in the star at those densities. We only plot the angular momentum
at densities larger than $10^{-11}~{\rm g~cm^{-3}}$.}
\label{specangmom}
\end{figure}

Shown in Fig.~\ref{velfieldfig} are filled contour plots of the logarithm of
the density and of the magnitude of the velocity field, over-plotted with
the velocity vectors.  In both cases, the pink curve indicates densities
between $8\times10^{-9}~{\rm g~cm^{-3}}$ and $10^{-8}~{\rm g~cm^{-3}}$, a
level at which we found a velocity $2-8~{\rm km~s^{-1}}$ (see
Table~\ref{velocitytable}).  In Fig.~\ref{specangmom} we show the specific
angular momentum as a function of the logarithm of the density.  The highest
specific angular momenta ($>10^{20}~{\rm cm^2~s^{-1}}$) are found at
densities below $\sim10^{-9}~{\rm g~cm^{-3}}$.  This lower density material
is farther from the
centre of the star, where the highest velocities are found.  At densities of
$\sim10^{-8}~{\rm g~cm^{-3}}$ there are also some regions with negative
specific angular momentum.  This is caused by an apparent rotation in the
star's envelope at those densities.  This is vaguely apparent by looking at
the velocity vectors in Fig.~\ref{velfieldfig} to the lower left of the
centre of the star.  In conclusion the star is spun up by the first
interaction such that the velocity contrast between the companion and the
gas at the next periastron passage will be smaller and the strength of the
interaction at the start of the CE could be reduced.

\subsection{The accretion disc}
\label{ssec:accretion}

The jets that formed the Rotten Egg nebula must have been massive and fast.
Our aim is to determine whether the conditions necessary for launching such
jets can arise during the modeled interaction.

\begin{figure*}
\includegraphics[width=0.95\textwidth]{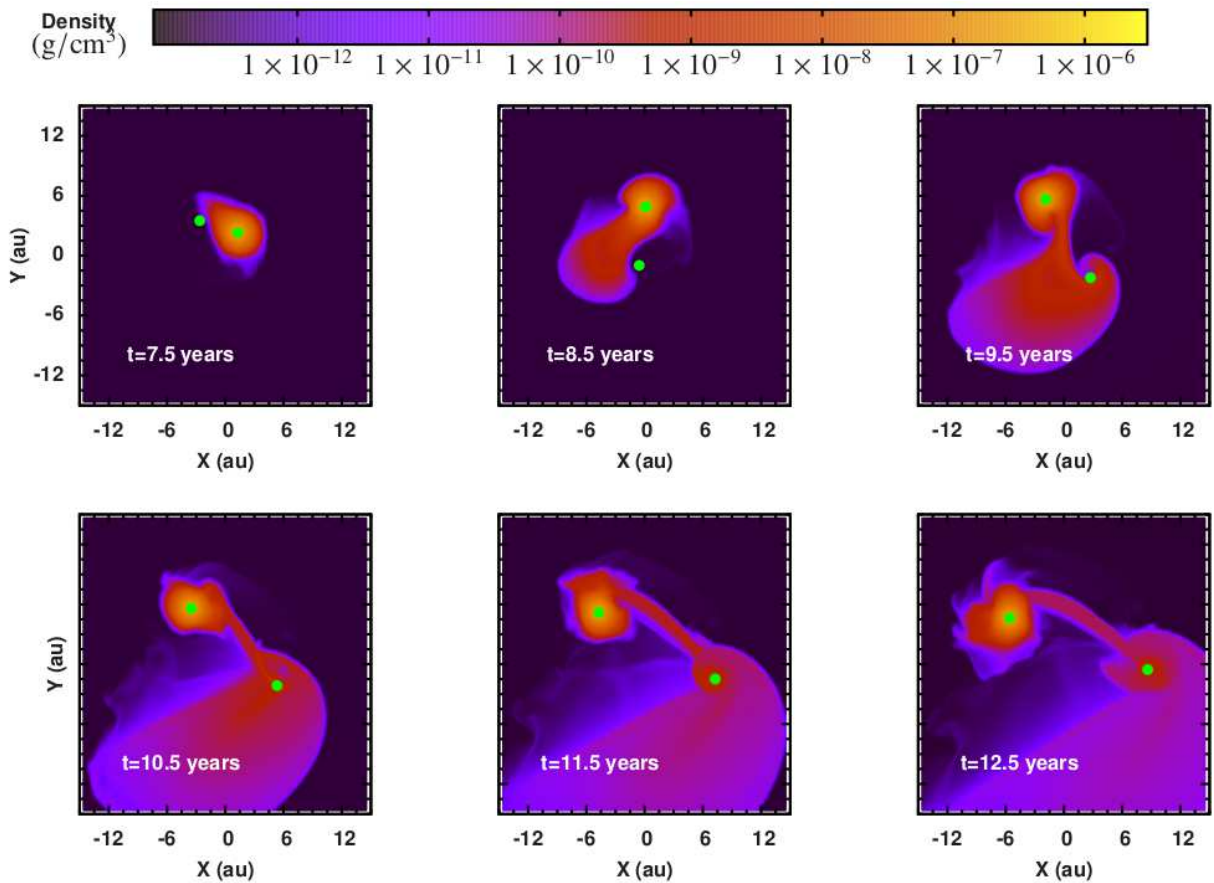}
\caption{Snapshots showing density slices in the equatorial plane around the
time of the first periastron passage for simulation \#1. The green dots are
the positions of the particles.}
\label{diskformation}
\end{figure*}

Here we discuss the formation of accretion discs at the time of RLOF, but 
acknowledge that once formed, the disc dynamics will be dictated by cooling and
magnetic fields, neither of which is included in our simulation.  We
therefore use primarily the mass of the disc after RLOF to anchor our
discussion, leaving all other conclusions to an analytical study in
Section~\ref{sec:jet}. There is an uncertainty in the disc mass since
some mass is lost from the grid following the first periastron passage.
However, the lost mass is substantially less than the mass of the
disc ($0.04~{\rm M_\odot}$) in simulation \#1, and most of it is
kinematically
unlikely to make it into the disc before the next periastron passage
(see Fig.~\ref{diskformation}).
Hence, the uncertainty in the disc mass following the first periastron
passage in simulation \#1 is small.

As we have discussed in Section~\ref{ssec:roche} we have adopted a high
value of the eccentricity to maximize the orbital period, while maintaining
a periastron distance that would allow mass transfer.  Hence the longest our
discs have to launch a jet after they form is 55 years (simulation \#6) or
27 years for a simulation with a more realistic initial eccentricity
(simulation \#2), although those discs contained very little mass.  The
longest time for a disc containing a reasonable amount of mass
($\gtrsim0.01$~\msun) to launch jets is 22~years (simulation \#2 after the
second periastron passage).  All these times should be regarded as upper
limits since for lower, more reasonable values of the eccentricity, these
times would be shorter.

\begin{figure}
\includegraphics[width=0.49\textwidth]{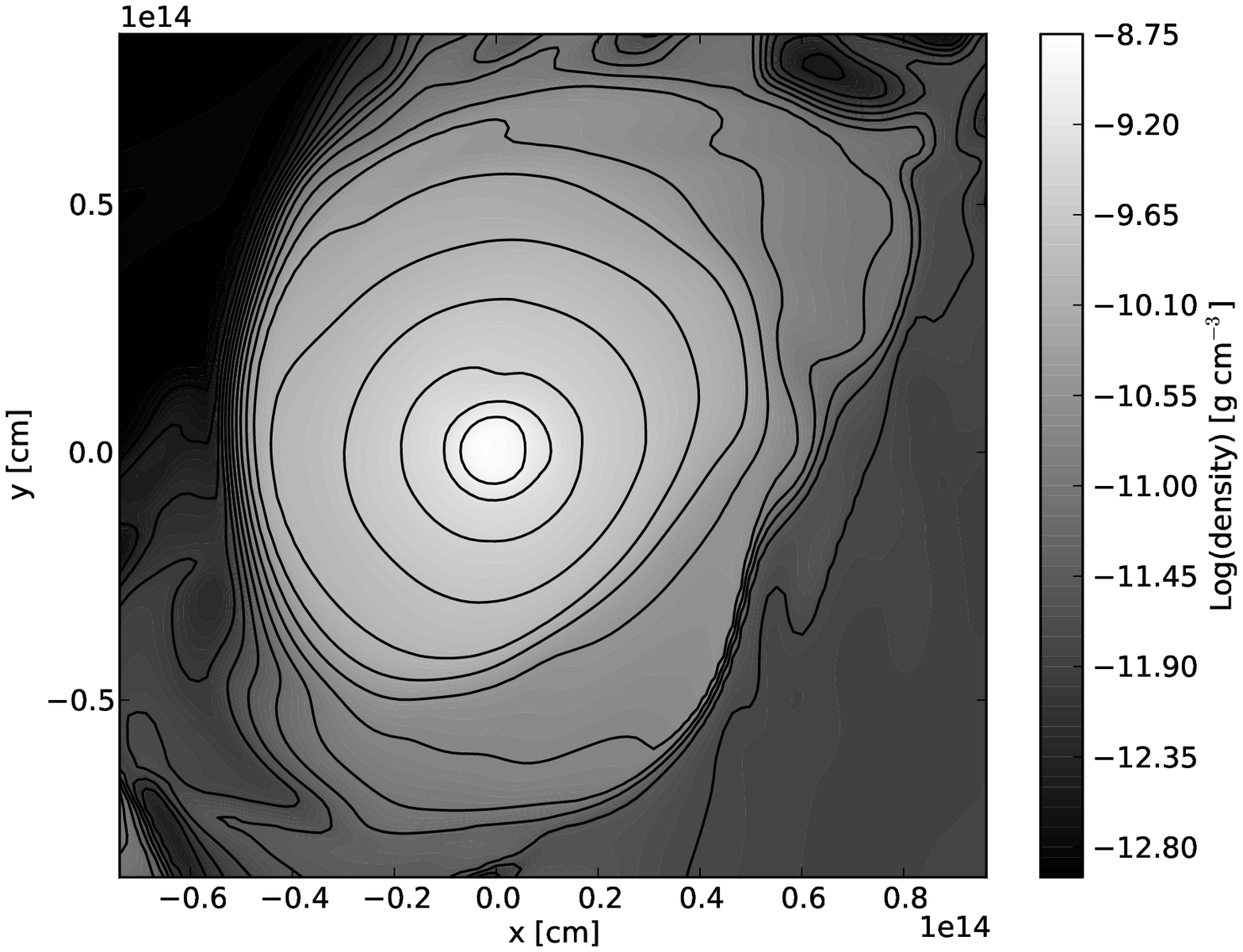}
\includegraphics[width=0.49\textwidth]{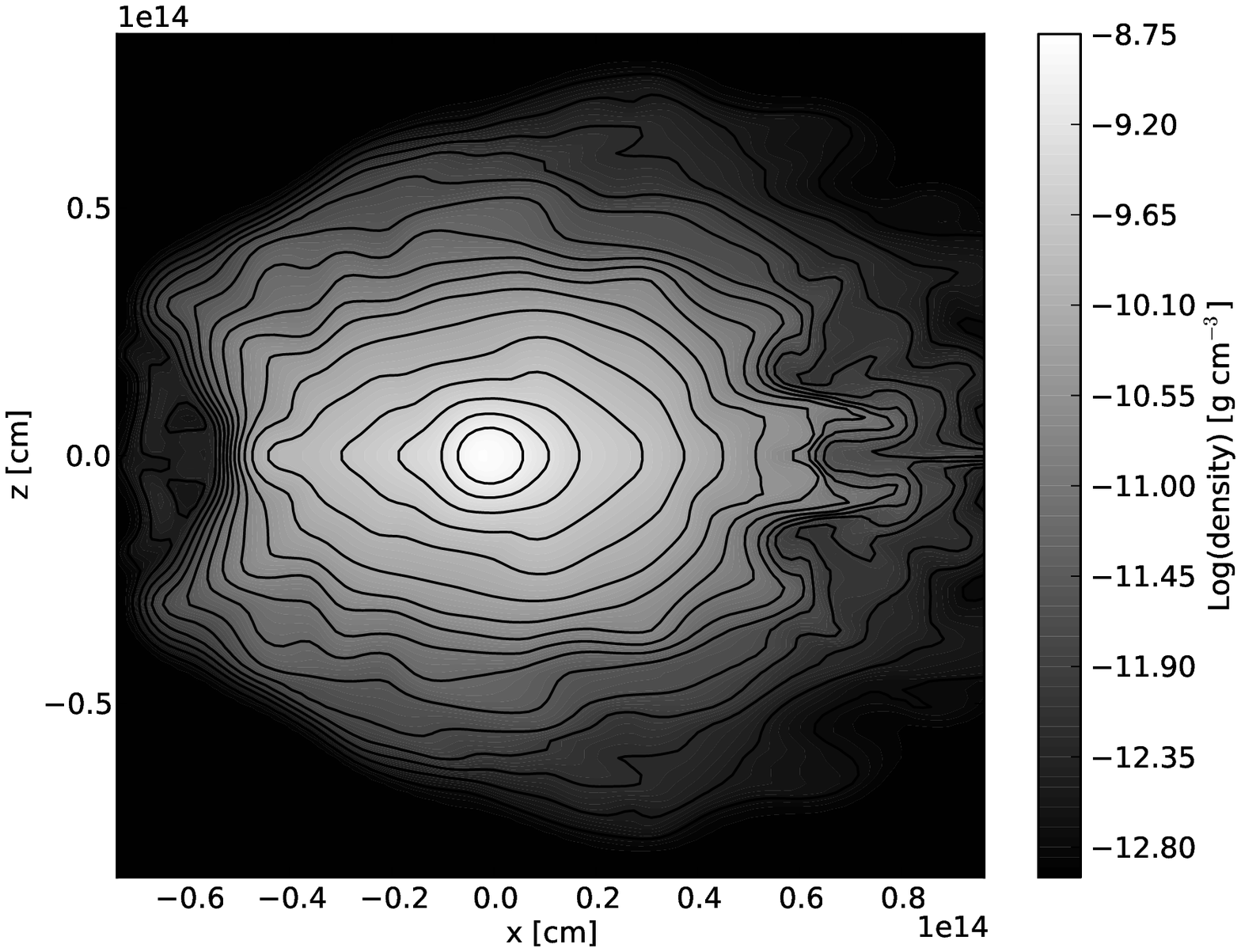}
\caption{A density slice in the disc's midplane (upper panel),
and in the perpendicular plane (lower panel) 12 years after the first periastron
passage.}
\label{discdensity}
\end{figure}

The masses of the discs formed in our simulations are $<0.06~{\rm M_\odot}$. 
The disc density structure is shown in Fig.~\ref{discdensity} for simulation
\#1.  The orbital period within that disc is 0.5 years at a radius of
$5\times10^{12}~{\rm cm}$ or 0.33~au, or 3 cells from the companion
particle.  Therefore in the $\sim10$~years available to form a
jet, the disc at that radius completes 20 revolutions, and many more at the
smaller radii close to the stellar surface from which a fast jet will be
driven \citep[e.g.][]{zanni13,staff15}.

We can gauge the mass accretion rate onto the disc from the following
argument.  When the companion approaches the first periastron passage, the
AGB star gets severely distorted and mass transfers to the
companion.  However, it is only some two years later that this mass begins to
engulf the companion, which has now passed periastron (see
Fig.~\ref{diskformation}).  Four to five years past first periastron
passage, one can start recognizing a disc around the companion, which is
linked to the AGB star through a stream of gas.  As the companion moves away
from the AGB star, the disc becomes more prominent while the stream weakens. 
Most of the $0.04~{\rm M_\odot}$ transferred to the disc arrives shortly
after periastron passage.  The mass transfer to the disc
varies with a maximum value of $\sim0.01~{\rm M_\odot~yr^{-1}}$
during this time.

The simulation with $2~{\rm R_{AGB}}$ periastron separation was also run with
eccentricities of 0.33, 0.5, and 0.9 (simulations \#4, \#5, and \#6 in
Table~\ref{initialsetups}).  A lower eccentricity results in a higher disc mass after the first periastron passage. This is relatively easy to understand: a lower eccentricity for a given periastron separation means a shorter period. Relatively more time is spent at close proximity for the more circular orbit. In simulation \#4, the lowest eccentricity simulated, the system almost forms a CE after
the first periastron passage. 

 \subsection{The effect of varying the companion mass}

In most of our simulations, the companion mass was $1.7~{\rm
M_\odot}$, because the companion in the Rotten
Egg Nebula is an A0V star.  However, since we find that upon mass transfer,
a CE interaction always follows within a few additional orbits, it
is difficult to explain how the companion can now be at a separation
$>10~{\rm au}$.  Other companion masses are therefore possible, since the A0V
companion may not have been responsible for the interaction, but instead a
third object may have interacted with the AGB star and merged.  This third
companion must have been less massive than the AGB star's initial mass of
$3.5~{\rm M_\odot}$, or otherwise it would have evolved sooner than the AGB
star.  We therefore ran one simulation with a companion mass of $0.6~{\rm
M_\odot}$ (simulation \#7), and another with a companion mass of $3~{\rm
M_\odot}$ (simulation \#8), both with an eccentricity of 0.7.  

A lower companion mass means that the periastron separation must be smaller
in order for any significant mass transfer to occur.  For equal
eccentricity, the orbital period will be shorter and the system enters a
common envelope phase sooner (when compared to simulation \#1).  For a
larger companion mass, a larger initial periastron separation leads to a
similar mass transfer, but with a longer period following the first
periastron passage.  Interestingly, both a smaller and a larger companion
mass than $1.7~{\rm M_\odot}$ lead to less mass being unbound from the
system (see Table~\ref{initialsetups}).  The larger final separation in
simulation \#8 (see below) may explain why less mass was unbound in the
simulation with the more massive companion.

\subsection{Orbital Stability and the Evolution of the Orbital Separation}
\label{ssec:separation}

\begin{figure}
\includegraphics[width=0.48\textwidth]{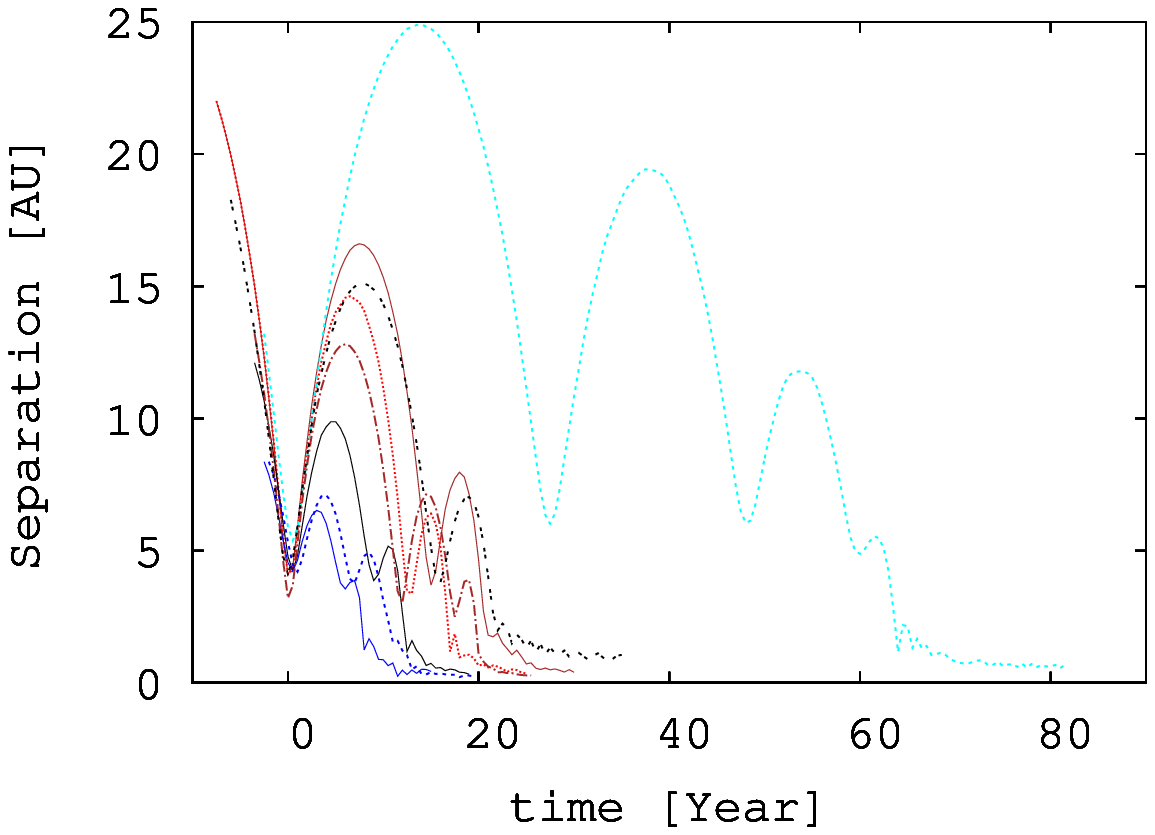}
\includegraphics[width=0.48\textwidth]{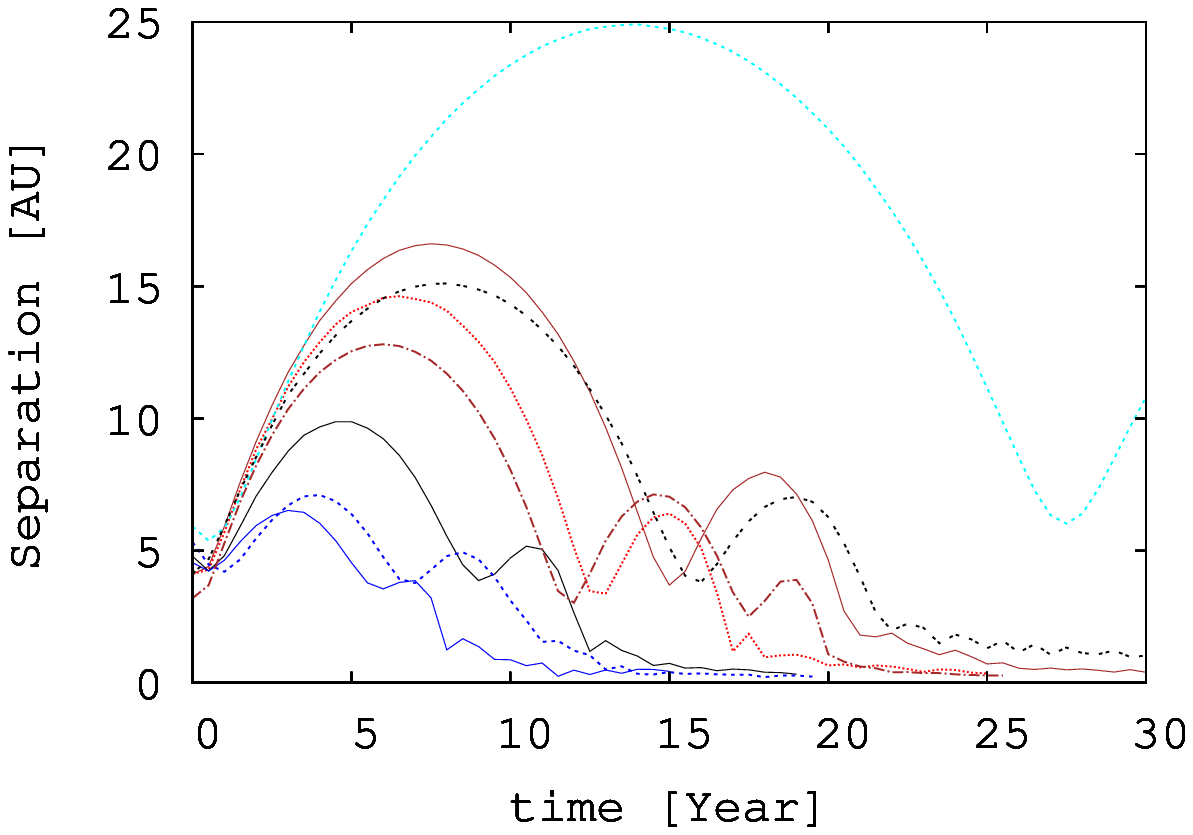}
\includegraphics[width=0.48\textwidth]{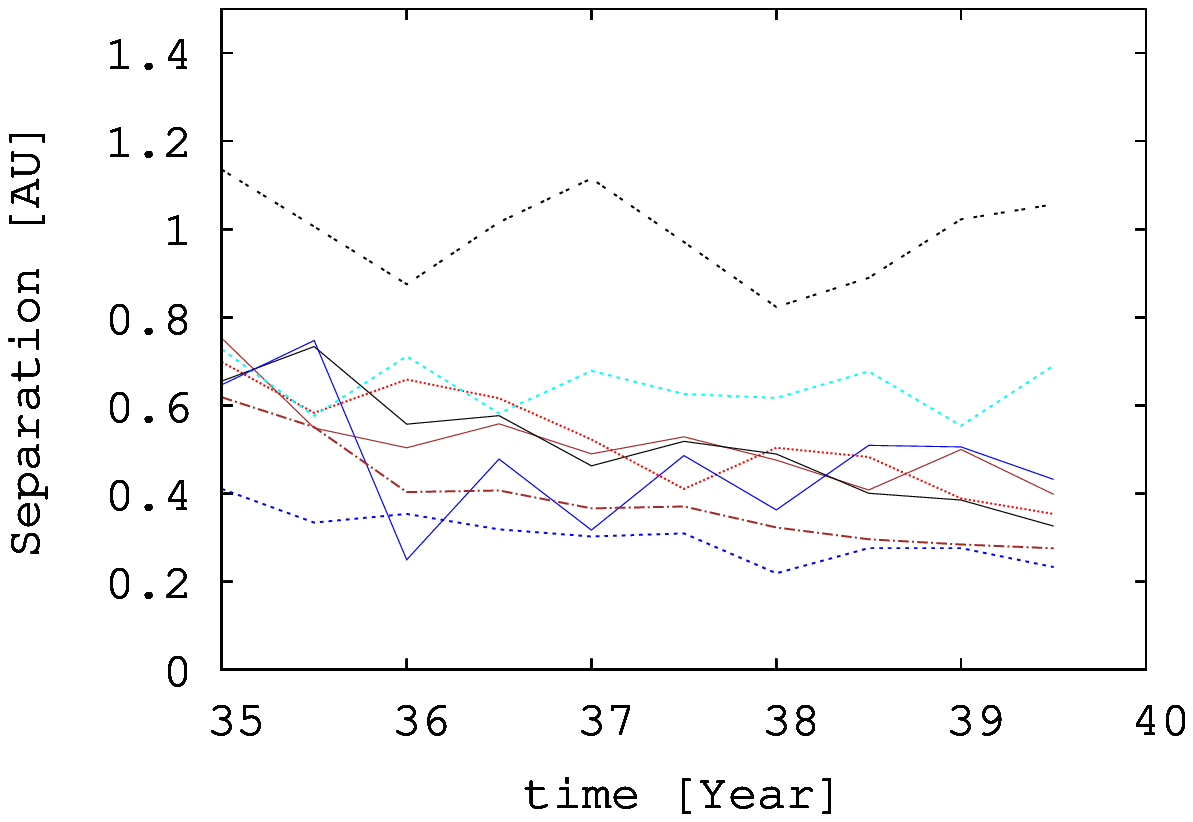}
\caption{{\it Upper panel:} The separation between the core of the AGB star 
and the companion as a function of time. Each curve has been shifted
so that the zero-time point coincides with the first periastron passage.
{\it Middle panel:} Zoom-in on the top panel between 0 and 30 years, to
better show the details of that part of the figure.
{\it Lower panel:} Zoom-in on the final separation between the core of the AGB
star and the companion. In this plot, the curves have been
shifted left or right so as to end at the same time. {\it Dotted red 
line:} simulation \#1ss (with 1.5 cells
smoothing length), {\it brown solid line:} 
simulation \#1 (with 3 cells smoothing length), {\it blue solid line:} 
simulation
\#4, {\it blue dashed line:} simulation \#4hr, {\it black line:} simulation 
\#5, {\it short-dashed cyan line:} simulation \#2, {\t brown dash-dotted 
line:} simulation \#7, {\it black dash-dashed line:} simulation \#8 
(see Table~\ref{initialsetups}).}
\label{35sep}
\end{figure}

\subsubsection{Early Evolution}

Simulations were also run with different periastron separations of
$2.5~{\rm R_{AGB}}$ and $3~{\rm R_{AGB}}$ (simulations \#2 and \#3 in
Table~\ref{initialsetups}).  In simulation \#3 the AGB star only receives a
small perturbation during periastron passage, but no mass is transferred.  The
simulation box is too small to capture the whole orbit, but we
could estimate the orbit based on the velocity vector before the companion
leaves the box after the periastron passage.  As it turns out, the orbit is
barely altered by the first periastron passage.  For simulation \#2 there
was a small amount of mass transfer during the first periastron passage,
which resulted in a $4\times10^{-4}~{\rm M_\odot}$ disc surrounding the companion. 
The orbit was altered and the orbital period and eccentricity changed from
36 to 27 years and from 0.7 to 0.65, respectively.  During the second
periastron passage 0.02~\msun\ were transferred to the disc and the orbit
was altered further (period and eccentricity changed to 22 years and 0.58,
respectively), eventually leading to a CE after four periastron passages.

Based on these simulations we conclude that a periastron separation of
roughly $2.5~{\rm R_{AGB}}$ is a critical separation for a companion mass of
$1.7~{\rm M_\odot}$.  For periastron
separations larger than this value the orbit is stable over longer, tidal
timescales.  For smaller separations, we observe instead a stronger
interaction that likely results in outburst type variability and leads
to a CE after one or two progressively smaller orbits.  In some sense,
increasing the periastron separation while maintaining the eccentricity
unchanged has a similar effect to increasing the eccentricity while
maintaining the initial periastron separation unchanged.  Less mass is
transferred during the periastron passage, and the system takes longer to
evolve into a CE.

For the simulations with more than one periastron passage, we find that the
eccentricity decreases for each passage (most noticeable for simulation
\#2 in Fig.~\ref{35sep}), as expected \citep{passy12}.  

\subsubsection{Final Separation}

In Fig.~\ref{35sep} we plot the separation between the core of the AGB star
and the companion as a function of time for all simulations that enter a CE. 
Initially, the separation is reduced, as the objects approach periastron. 
The final separations are about 0.4 au for all of simulations, except
simulation \#2, which has a larger final separation of $\sim$0.6 au, the
higher resolution simulation \#4hr with a final separation of $\sim0.2~{\rm
au}$, and simulation \#8 with a final separation of $\sim 1~{\rm au}$.  
These separations are only upper limits, however, because the gravitational
smoothing length of
3 cells corresponds to 0.34~au for the 256$^3$ resolution (simulation \#1
and \#7)
and 0.17~au for the other $256^3$ simulations with 1.5 cells smoothing
length, as well as the $512^3$ simulation with 3 cells smoothing length
(simulation \#4hr). Comparing simulation \#1 to simulation \#1ss shows that
a factor of two decrease in smoothing length also does not change the result,
although both values are within a factor of two of the final orbital radius.
However, the comparison of simulation \#4 to simulation \#4hr shows that
increasing the numerical resolution by a factor of two, even with constant
smoothing length, causes a factor of two decrease in the final orbital
radius. This suggests that resolution of the orbit rather than smoothing
length is the dominant factor determining final radius \citep[also
see][]{passy12}. The final radius reached with these physical assumptions
will ultimately need to be determined by adaptive mesh simulations.

We do note that the stabilization separation is larger (0.6~au) for the
simulation that starts with more angular momentum (simulation \#2).
We attribute this trend to the fact that more angular
momentum is transferred to the primary, which is spun up as a result.  This
has the effect of lessening the strength of the interaction with the
companion, although
this is still a poorly resolved result.

\subsection{Mass loss}
\label{ssec:massloss}

\begin{figure}
\includegraphics[width=0.48\textwidth]{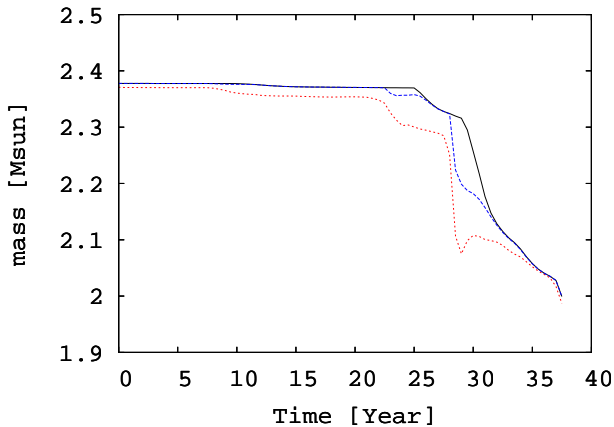}
\caption{{\it Solid black curve:} The total gas mass on the grid (i.e., excluding the 
mass of the particles) for simulation \#1. {\it Dotted red curve:} the total bound gas mass on the
grid, including the thermal energy in the calculation of unbound mass. 
{\it Dashed blue curve:} the total bound gas mass on the grid, excluding the thermal
energy calculating the unbound mass.}
\label{boundmass}
\end{figure}

\begin{figure}
\includegraphics[width=0.48\textwidth]{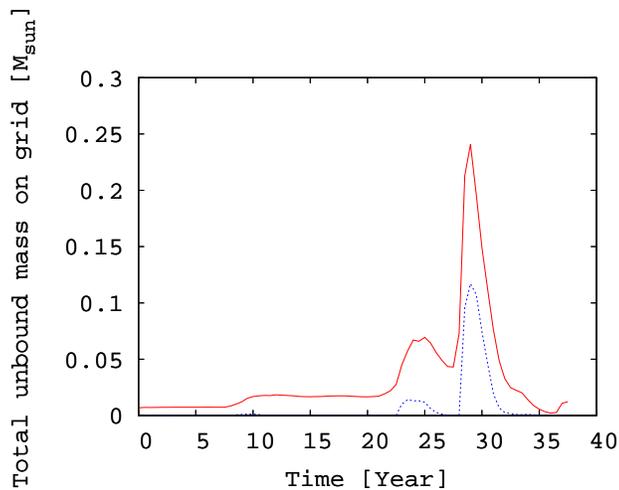}
\caption{The total unbound mass on the grid as a function of time for
simulation \#1, including 
the thermal energy term when calculating the unbound mass (solid red curve) or 
excluding it (dashed blue curve). Rises in the curves correspond to unbinding 
events, while drops occur when unbound mass flows off the grid (cf. 
Fig.~\ref{boundmass}).}
\label{unboundmass}
\end{figure}

An outstanding question in all binary interactions, including mergers, is
how much mass is lost from the system.  In CE interactions resulting in
close binaries we know that the entire envelope of the giant star must be
ejected.  However, simulations \citep{rasio96,sandquist98,ricker12,passy12} have
typically found that most envelope mass is still bound to the system by
the time the orbital in-fall stabilizes, opening the possibility that either
gas falls back rekindling the interaction, or pointing to the fact that an
additional energy source must be available \citep{Ivanova2015,nandez15}.  

Because of the limited size of the grid, mass flows out of the computational
domain, especially towards the end of the simulations.  Fig.~\ref{boundmass}
shows the total gas mass on the grid (i.e., excluding the mass of the
particles) and the bound gas mass on the grid for simulation \#1.  As can be
seen, most of the mass remains bound.  In Fig.~\ref{unboundmass} we show the
total unbound gas mass on the grid as a function of time for simulation \#1. 

Both the bound and the unbound mass shown in Figs.~\ref{boundmass} and
\ref{unboundmass} have been calculated both including and excluding the
thermal energy term.  After the first periastron passage ($\sim8$ years), we
find that about $0.02~{\rm M_\odot}$ is unbound when including the thermal
energy, and only a negligible amount when excluding it.  After the second
periastron passage, the unbound mass reaches about $0.07~{\rm M_\odot}$
($0.01~{\rm M_\odot}$, excluding the thermal energy) before dropping by
about $0.03~{\rm M_\odot}$ ($0.01~{\rm M_\odot}$), as some unbound mass
leaves the computational domain.  Then the unbound mass grows by about
$0.21~{\rm M_\odot}$ ($0.10~{\rm M_\odot}$) to reach a peak of about
$0.25~{\rm M_\odot}$ ($0.12~{\rm M_\odot}$).  Summing up, $0.28~{\rm
M_\odot}$ ($0.13~{\rm M_\odot}$) of gas in total was unbound, or just above
10 per cent (5 per cent) of the AGB star's envelope.  In
Fig.~\ref{unboundmass} a decreasing
unbound mass is caused by unbound
mass leaving the grid or by unbound mass losing energy and becoming
bound, for instance by colliding with bound of gas.  Following the
final periastron passage, much mass, both bound and unbound, flows off
the grid.  Eventually the unbound mass drops to zero, indicating that no
more mass is being unbound as the separation stabilizes.  Since the final
separation is an upper limit, we must conclude that the unbound mass in our
simulations is a lower limit.

\subsection{Numerical considerations}
\label{ssec:energy}

\subsubsection{Energy Conservation}

\begin{figure*}
\input{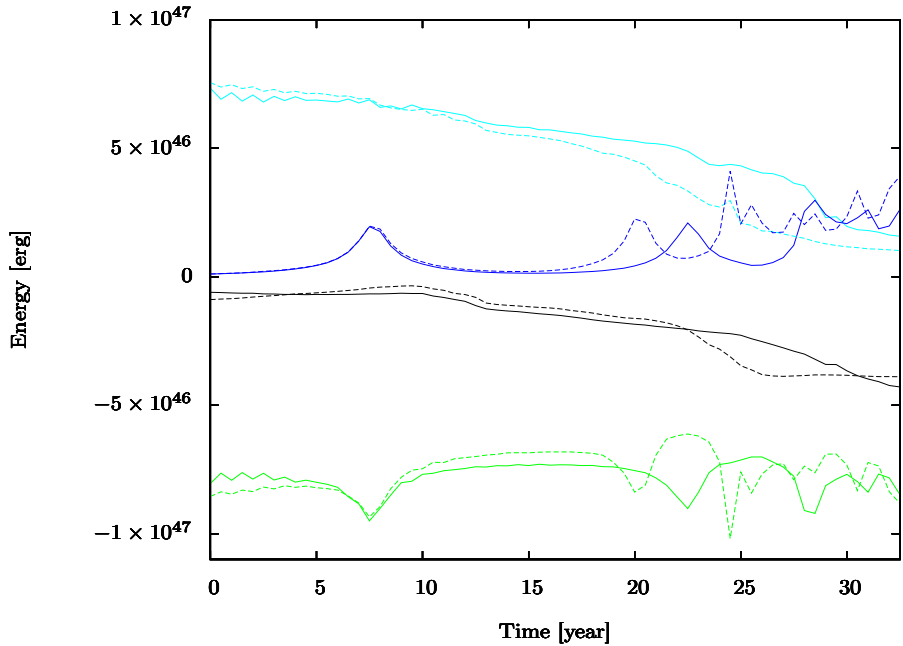}
\caption{Energy components for simulation \#1 (solid curves) and simulation
\#1ss (dashed curves) (3 cells and 1.5 cells smoothing length respectively). 
The blue
curve is the kinetic energy of the gas in the AGB star's envelope and of the
particles, the cyan curve is the
thermal energy in the box, the green curve is the total potential
energy of the gas-gas, gas-particles, and particle-particle,
and the black curve is the
total energy (kinetic+potential+thermal) of the whole system in the
computational domain.}
\label{energies2rs10}
\end{figure*}

No grid
simulation of this kind conserves energy perfectly \citep[see,
e.g..][]{sandquist98,passy12}. This is in part due to the contribution from
self-gravity being added as a source term in the momentum and the energy
equations \citep[Eqs. 2 and 3 in][]{bryan14}. In other words, these
equations are written and solved in non-conservative form. Therefore,
momentum and energy are not conserved to roundoff error and the level of
non-conservation can vary depending on how dominant self-gravity is for the
evolution of the system. For more details, see \citet{jiang13}.
The steeper the potential gradient, the worse the conservation. 

Checking for energy conservation is notoriously difficult in grid
simulations with outflowing boundary conditions because mass and energy are 
lost from
the computational domain.  This situation is particularly difficult to
handle in our case.  Our low density, high temperature ``vacuum'' (see
Sec.~\ref{sec:method}) has a
substantial amount of thermal energy, which leaves the grid even before
substantial mass starts to flow out.  \citet{passy12} used their SPH
calculation to determine that energy and angular momentum were reasonably
conserved.  They did not check their grid simulations for energy
conservation because the results of grid and SPH simulations, for the same
input parameters, were comparable. 

In Figure~\ref{energies2rs10} we show the energy components of simulation
\#1 compared to \#1ss (which has a smaller smoothing length of 1.5 cells,
for which energy conservation tends to be worse).  The thermal energy
(cyan curve) is dominated to large extent by the hot, low density gas
initially permeating the computational domain \citep[which we call
vacuum;][]{passy12}.  This low density medium contains substantial thermal
energy despite its low total mass.  In addition, some of the AGB star's
thermal energy is converted into potential energy as the star expands,
adding to the decline in the thermal energy.  By the time of the first
periastron passage at $t\approx8~{\rm years}$, when substantial mass starts
flowing out of the computational domain, making it difficult to track energy
conservation, the simulation box has lost $2\times10^{-4}~{\rm M_\odot}$ of
this hot, low density ambient gas.  The thermal energy of this lost gas is
about $10^{45}~{\rm erg}$ for both simulations, explaining the initial drop
in the thermal energy.

Despite the drop in thermal energy, the total energy in simulation \#1 shows 
a slight increase
over a four year period around the first periastron
passage.
This increase is about $5\times10^{44}$ erg over these four years, 
and because it is small compared to the scale of the plot in
Fig.~\ref{energies2rs10} it is not apparent in the figure. During this
time, the total energy should in fact have decreased by about
$5\times10^{44}$
erg because of the hot, low density ambient gas left the grid. 
The error over this four year period is therefore $\sim1\times10^{45}$
erg, which we can compare to the total initial potential energy\footnote{We
compare the change in total energy to the initial stellar potential energy
rather than the initial total energy in the computational domain.  The
stellar potential energy can be considered representative of the stellar
energy because the kinetic and internal contributions are small (the thermal
energy in the computational domain is dominated by the energy of the hot
vacuum).} of $\sim8\times10^{46}$ erg.  The relative error in the total
energy over this four year period is therefore $\sim1\%$.  In simulation
\#1ss, with a smaller smoothing length of 1.5 cells, we find a larger error
in the energy conservation of $4\%$ over the same period.  We note however
that energy non-conservation in simulation \#1 is only apparent during the 
4 years around the
first periastron passage, when the interaction is strongest. We therefore do
not expect this non-conservation to grow linearly with time, at least not
for the period of time between 10 and 20 years, when the primary core and
companion are not interacting strongly. Indeed, for the first 9 years of
simulation \#1, we find that the error in the energy conservation is also
$\sim1\%$, while in simulation \#1ss it is $\sim8\%$ over the first ten
years.

From approximately 10 years after the start of the simulations, sufficient
mass leaves the grid that energy conservation cannot be tracked any longer. 
We note that in simulation \#1ss, with the shorter smoothing length, the
evolution proceeds faster as expected from the fact that the gravitational
interaction between particles and gas is stronger.  Although the energy
conservation is worse, the strength of the interaction is more realistic. 
Overall, however, the outcome of the simulations are very similar, when
considering our tolerance.  In simulation \#1ss the second periastron
passage takes place $\sim$2 years earlier and the disc mass is half that of
simulation \#1.  The amount of unbound gas and the post-CE separations are
very similar.

\subsubsection{Numerical resolution}
\label{sssec:resolution}

We carried out a simulation (\#4hr) equivalent to simulation \#4, but with a
resolution of 512$^3$ instead of 256$^3$.  The smoothing length of
simulation \#4hr was 3 cells, compared to the 1.5 cells of simulation \#4. 
In this way the strength of the interaction between particles and gas is
preserved.

The higher resolution simulation~\#4hr has a slightly weaker interaction at
periastron, causing slightly less mass to be transferred during the first
periastron passage and therefore a slightly longer orbital period following
that (see Table~\ref{initialsetups}).  Simulation~\#4hr unbinds much less
mass than simulation~\#4 (0.16 vs.  0.42~\msun) if considering the thermal
energy in deciding what mass is unbound, however, if only considering
whether gas has reached the escape velocity when determining if mass is
unbound, the difference is found to be less (0.12 vs.  0.15~\msun).  The
lower final separation in simulation \#4hr (0.2 vs.  0.4~au) indicates that
the resolution affects the final separation, as expected \citep{passy12} and
discussed above.

However, the exact role played by the combination of resolution and
smoothing length in slowing down the in-spiral is not simple.  Within a
particle's smoothing length, the gradient of the gravitational potential is
zero.  This therefore affects how close the particles can reliably be
simulated.  For both simulations \#4 and \#4hr, the respective smoothing
lengths are equivalent to $2.6\times10^{12}~{\rm cm}=0.17~{\rm au}$, only
slightly smaller than the 0.2~au separation reached by simulation \#4hr.  We
also know from comparing simulations \#1 and \#1ss that the smoothing length
per se did not affect the final separation.  It may be that the larger
growth in potential energy in simulation \#1ss than in simulation \#1 due to
energy non-conservation leaves the companion at a larger separation than
what would have resulted had the energy been perfectly conserved.

Pending additional tests of how the gravitational friction that dictates the
extent of the inspiral depends on different numerical aspects, we can only
say that the simulation is not formally converged.  This said, and in light
of the other tests we have carried out and of the fact that we are, at this
time, interested primarily in an order of magnitude answer, we did not carry
out additional convergence testing.

\subsubsection{The adiabatic assumption}
\label{sssec:adiabatic}

Our simulation is carried out assuming an adiabatic gas. This is appropriate
when simulating the fast, dynamical inspiral phase of a CE interaction, but
less so for the current simulations that take place over 20-30 years, which is 6-10 AGB 
star dynamical times and comparable to the $\sim$45 year AGB star thermal timescale.  We
here make a rough estimate of the energy that would be lost due to
radiation, to estimate how appropriate the adiabatic assumption is in
our case.

Following the first periastron passage in simulation \#1, the AGB star
approximately doubles its radius.  If we assume that it retains the same
surface temperature, then we find that during the 14 years between the first
and the second periastron passages the star would radiate $10^{47}~{\rm
ergs}$.  This is a factor 4 larger than the energy radiated by the
unperturbed star over this 14 year time period ($2.5\times10^{46}~{\rm
ergs}$).  The thermal energy of the unperturbed star initially is
$4\times10^{46}~{\rm ergs}$.  Hence the star does not have sufficient energy
to maintain this luminosity for the whole period between the two first
periastron passages.  In reality the photospheric temperature of the
expanded AGB star may be lower than its initial effective temperature. 
Considering as a lower limit that the luminosity of the expanding AGB star
remains the same, its photospheric temperature must decrease to
$\sim2060~{\rm K}$ (from the initial temperature of $2920~{\rm K}$).  Mira
stars have a pulsation cycle of approximately a year with a radius excursion
of a factor of $\sim$2, during which time the photospheric temperature
decreases by approximately the same factor \citep{Ireland2011}.  If this were
the case in our situation the luminosity of the expanded AGB star would be
approximately one tenth that of the original star.  We conclude that the
adiabatic assumption is likely a borderline assumption in this case and may,
during the second periastron
passage, have contributed somewhat to a larger AGB star than would result
if energy losses were included.

The disc forming around the companion after the first periastron passage in
simulation \#1 has a mass of $0.04~{\rm M_\odot}$, but it has a radius of
several au.  However, with such low mass, the disc possesses little thermal
energy compared to the star, and it is therefore likely that its photosphere
would cool quickly.  Hence, once again we may be overestimating the
interaction during the second periastron passage since both the AGB star and
the disc may be thinner than we simulated.  However, since the periastron
separation decreases between the first two passages (see Fig.~\ref{35sep}),
we conclude that although the simulated interaction at the second passage
might be stronger than in nature, so the number of orbital periods before
the CE phase may be lower, it is unlikely to be lengthened
substantially.

\section{Possible jet parameters}
\label{sec:jet}

The outflows in the Rotten Egg nebula have velocities of $200 - 400~{\rm
km~s^{-1}}$ \citep{desmurs12}.  \citet{sanchezcontreras04} suggested that
these lobes had been inflated by jets with velocities of the order
$500-1000~{\rm km~s^{-1}}$. \citet{soker12} found that, to be consistent with the energy and
momentum measurements of  \citet{alcolea2001}, the jets must have had a 
mass of $0.01-0.02~{\rm M_\odot}$.  The jets then interacted with the circumbinary
gas to create the present-day lobes with their slower velocities.  

If we assume that the jets originate in a disc around the companion, and that
the outflow velocity equals the escape velocity ($v_{\rm esc}=\sqrt{2GM/r}$,
with $M=1.7~{\rm M_\odot}$ being the mass of the companion and $r$ being the
distance from it) at the launch radius, then we find that the jet launching 
radius must be $2.6$~\rsun\ to reach a velocity of $500~{\rm km~s^{-1}}$. 
The lower limit for the disc launch radius is the stellar radius, $\sim$1.5~\rsun\ for a
$1.7~{\rm M_\odot}$ main sequence star, and the maximum jet speed achievable is $\sim655~{\rm
km~s^{-1}}$.

The maximum accretion rate into the disc at the time of RLOF in our simulations is
0.01~\msunyr\ (Sec.~\ref{ssec:accretion}).  If this were the accretion rate
onto the companion it would be entirely plausible for a jet of 0.02~\msun\
to be launched over a decade, assuming that 10-40\% of
the accreted mass is launched into a jet \citep[such ratios are typical for
protostellar jets, see for instance][]{pudritz12,federrath14}.  However, such an accretion rate onto the
secondary would generate a luminosity above the Eddington limit. 
The Eddington luminosity

\begin{equation}
L_{\rm Edd} = \frac{4\pi GMm_{\rm p}c}{\sigma_{\rm T}} = 
5.4\times10^4 L_\odot \times \bigg(\frac{M}{1.7 M_\odot}\bigg), 
\end{equation}

\noindent where $m_{\rm p}$ is the proton mass and $\sigma_{\rm T}$ is the
Thomson scattering cross section. This
translates into an upper limit for the accretion rate of

\begin{dmath}
\dot{M} = (0.002-0.003 \ M_\odot {\rm yr}^{-1})  \bigg( \frac{L}{54\,000~L_\odot} \bigg) \times \\
\bigg(\frac{r}{2.6-1.5~R_\odot}\bigg) \bigg( \frac{1.7 M_\odot}{M} \bigg) , 
\end{dmath}

\noindent Assuming that 10-40\% of the accreted mass is launched into a jet,
we find that in ten years a jet of no more than $0.2-1.2\times10^{-2}~{\rm
M_\odot}$ could be launched, a range that still overlaps with the lower
range of the estimates of \citet{soker12}.  However, the Eddington
luminosity calculation assumes spherical accretion.  The disc accretion
considered here allows greater accretion rates
\citep[e.g.,][]{krumholz2009,shiber15}, something that we discuss further in
Sec.~\ref{ssec:discussiondisk}.

\begin{figure}
\includegraphics[width=0.48\textwidth]{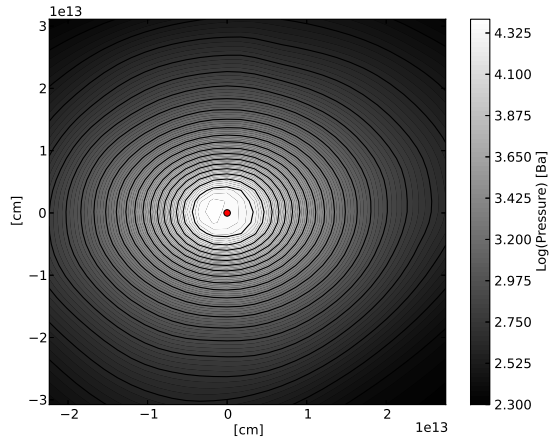}
\caption{Contour plot showing the logarithm of the pressure in the disc
midplane around the companion, 20 years into simulation \#1. The position of
the companion is indicated by a red circle.}
\label{diskpressure}
\end{figure}

Next, we determine the magnetic field needed to launch the jets to check its consistency 
with the calculations above. We assume that the jet is launched by
a magnetic field threading the disc. To constrain the magnetic field strength we assume equipartition between the gas
pressure in the disc and the magnetic pressure, $P_{\rm m}=B^2/8\pi$.  The gas
pressure close to the companion is shown in Fig.~\ref{diskpressure}. This
pressure value must be a lower limit because our grid resolution is an order
of magnitude too large to resolve the inner part of the disc. Therefore we
find that close to the companion the gas pressure is $>10^{4.3}~{\rm Ba}$
and that the magnetic field must therefore be $>700~{\rm G}$.

\citet{wardle07} and \citet{tocknell14} estimated the magnetic field 
 in an accreting disc using a formalism designed originally for accretion
discs around forming stars.  They assumed that the magnetic field
responsible for the jet also produces the viscosity that leads to angular
momentum loss and accretion (if another form of viscosity were to be present
this would overestimate the magnetic field strength).  We note that this
expression is {\it local} to a given disc radius and does not attempt to
describe how the magnetic field strength redistributes in response to
accretion.  We also note that this expression is for a geometrically thin
disc.  However, the difference for a thicker disc would be of the order of a
factor of $\sqrt{r/H}$ (where H is the pressure scale height), which for us
is only a factor of 1.2, even neglecting the effect of disc cooling, which
will lead to a thinner disc.

Following \citet{wardle07} and using as the local
accretion rate $\dot{M}=0.003$~\msunyr\ at radius $r$=1.5~\rsun, the poloidal 
magnetic field at that location can be found to be:
\begin{dmath}
B_p\approx(19~{\rm kG}) 
\bigg(\frac{\dot{M}}{0.003~{\rm M_\odot~yr^{-1}}}\bigg)^{1/2} \times \\
\bigg(\frac{r}{1.5~R_\odot}\bigg)^{-5/4} 
\bigg(\frac{M}{1.7~{\rm M}_\odot}\bigg)^{1/4}.
\label{accrateeq}
\end{dmath}

\noindent With this poloidal field, the toroidal component will
be a factor of several larger, depending on the value assumed for the viscous
$\alpha$ parameter \citep{Shakura73,blackman01}.  

The radial dependence of the gas pressure in the disc is well matched by a power-law: $P\propto
r^{-2}$ for distances from the companion larger than $\sim73~{\rm R_\odot}$, corresponding to the 3
cell smoothing length.  Extrapolating this power law to
$1.5$~\rsun\ from the companion, we find that the pressure there would be $\sim3.5\times 10^7~{\rm
Ba}$, equal to the magnetic pressure of a $30~{\rm kG}$ magnetic field. 
This value is in order-of-magnitude agreement with the magnetic field strength value
inferred using the viscosity argument of Wardle (2007).

 Since a lower companion mass requires a shorter orbit in order for
significant mass transfer to take place and a massive disc to form around
the companion, the accretion rate must be even larger than we considered
above in order to accrete the whole disc before the next periastron passage
and the following CE interaction.  A larger companion mass, on the other
hand, can cause greater mass transfer even with a larger periastron
separation.  In this case then, a longer orbit following the first
periastron passage is possible and it might be more likely that the whole
disc accretes before the next periastron passage occurs.  We also found the
most massive disc in our simulations in simulation \#8, so a more massive
companion appears most likely to lead to a massive jet.

\section{Discussion}
\label{sec:discussion}

\subsection{Orbital stability after Roche lobe overflow}
\label{ssec:discussionorbital}

 A substantial disc forms in our simulation with mass ratio $q=0.55$
only for $R<2.5 R_{\rm star}$ or $5.5~{\rm au}$. Larger values of $q$ allow for larger
separations, while smaller values of $q$ require smaller separations.
For a 3.05~\msun\ AGB star the maximum AGB
radius during thermal pulses is between 2.6 and 3.3~au \citep{madappatt15}.
Hence the periastron distance at which RLOF takes place could be
as far as 8~au.
For an interaction to happen at periastron passage, orbital periods cannot
exceed a couple of decades or else the eccentricity would be
unrealistically large.  In addition, if a sufficiently massive disc is to be
formed the orbit is shortened and the time to the inevitable CE becomes very
short indeed (2-3 orbital periods).  

For separations of $R\simeq2.5~R_{\rm AGB}$  (and $q=0.55$), a less massive
disc is formed but during the next periastron passage the situation becomes,
again, that described above, and a sufficiently strong interaction forms a disc
of a few hundredths of solar mass, after which a CE is inevitable.  For
larger periastron separations of $R\gtrsim R_{\rm AGB}$, a RLOF disc does
not form (although a much less massive disc can form from accretion of wind
material; \citealt{HuarteEspinosa2013}) and the orbit is stable over tidal
timescales.

\subsection{The Post-CE separation and unbound mass}
\label{ssec:discussionCE}

In the present simulations, at most 10 per cent of the AGB star's envelope
mass has been unbound during the CE interaction. The final separations of
the CE cores range between 0.2 and 0.6 au (Table~\ref{initialsetups}). If the final separation were
affected by the limited resolution, the final separations in our simulations
would be upper limits and the unbound masses would be lower limits. 

The question of what the post-CE separation should be remains wide open.
Simulations tend to predict larger, post-dynamical phase separations
compared to observations \citep[e.g.][]{passy12}, but the phase
immediately following the post-dynamical phase \citep[phases III an IV
in][]{ivanova13}
could alter that separation further by any of a number of
mechanisms. In-fall of bound material would reduce the separation further.
On the other hand, the action of energy sources unaccounted for in
current simulations could help eject the envelope contributing to larger
separations \citep[e.g.][]{nandez15}. In summary, we leave the question of
the final separations unanswered, including the issue of whether a system
like this could ever merge (see also Section 5.4). However, if a merger
does not occur, the final separation will be small, much less than 1 au. For
the case of the Rotten Egg Nebula the binary must have merged because the
primary is still an AGB star with an envelope.

\subsection{Disc and Jets}
\label{ssec:discussiondisk}

The main goal of this study was to determine how massive a disc 
would be formed via RLOF at periastron passage.  The accretion
discs in our simulations have masses of a few hundredths of a solar mass and last on the order of a
decade.  The peak mass accretion rate into the disc is
0.01~\msun~yr$^{-1}$.  This mass is distributed in a slightly flattened,
rotating structure with a large scale height of $\sim0.7\ r$.  The
orbital period is half a year at $\sim$64~\rsun.  How this disc changes
over the decade after formation and before more mass is added to it
or a CE destroys it, depends on the magnetic field and on the radiative 
properties of the gas.  

The peak accretion rate onto the disc, 0.01~${\rm M_\odot~yr^{-1}}$ would be sufficient to create a jet of
the needed mass over the available time only if that accretion rate were
maintained across the disc radius and onto the companion.  This accretion rate would be super-Eddington. Limiting ourselves to the Eddington accretion of $0.002-0.003~{\rm
M_\odot~yr^{-1}}$, could generate jets with a mass of only $0.2-1.2\times10^{-2}~{\rm M_\odot}$, which is at the lower end of the range of $1-2\times10^{-2}~{\rm M_\odot}$ discussed by \citet{soker12}. The momentum of our highest mass jet would also be too low since its velocity could only be as high as 655~km~s$^{-1}$, short of the needed 1000~km~s$^{-1}$.

This said, our jets are not much less massive, or less energetic than required by the observations and the error bar both in the observations and in our simulations may bring the values into agreement.
Additionally there are other possibilities that would boost the predicted jet mass: the
jet launch efficiency could be higher and/or
additional gas ejected
following the second periastron passage, when the simulations show that there is a brief period of
$\sim4-5$ years before the third periastron passage leads to the CE.  
Another, more likely possibility is that the accretion is super-Eddington. The Eddington limit applies to spherical accretion and assumes that all the generated energy goes into radiation. Neither of these assumptions likely applies in our situation, in which the accretion is likely equatorial and much of the energy goes into kinetic energy of the jets.

Finally we point out that if the companion had been a WD, escape speeds (and
hence jet speeds) as high
as 1000~km~s$^{-1}$ would easily have been achieved.  However, the mass
accretion rates would be lower than for a main sequence star, once the
combination of plausible masses and overall lower accretion radii are
combined with the Eddington limit reasoning above. A typical white
dwarf will have a mass of $\sim0.6~{\rm M_\odot}$ as in simulation \#7,
where we found that because of the companion's lower mass, the
periastron separation must be smaller in order to achieve significant mass
transfer, which means that the orbital period will be shorter. Hence, there
would be very little time available to launch a jet, necessitating very high
accretion rates in order to produce jets like those observed in the Rotten Egg
Nebula. 

A more massive main sequence companion (simulation \#8) can also allow for
higher ejection speeds, as the escape speed from main sequence stars
increases with mass.  However, since the companion cannot be much more
massive than
$3~{\rm M_\odot}$, this can only increase the escape speed from the object
by $\sim15-20\%$.  A more massive companion does
allow for a more massive disc and a longer period following the
first periastron passage, allowing for more mass to be launched into a jet,
but a merger is less likely \citep{DeMarco2011}.

\subsection{The status of the Rotten Egg Nebula}
\label{ssec:discussionOH231}

The central star in the Rotten Egg Nebula is observed to be a binary
containing an AGB star and a main sequence companion at a separation larger
than 10~au \citep{sanchezcontreras04}.  The nebula includes a relatively
massive circumbinary torus and
collimated lobes. In this paper we have scrutinized the proposal
\citep{soker12} that the
nebula formed $\sim$800 years ago, during a periastron passage of the
binary, by the mechanism of RLOF and jet formation.

To do so, we have performed a number of grid based hydrodynamic simulations
of the interaction
of a $3.05~{\rm M_\odot}$ AGB star and a less massive companion in an
eccentric orbit,
with a range of eccentricities, initial periastron separations and companion
masses.
We have found that for periastron separations
greater than $\sim 2.5~{\rm
R_{AGB}}\approx5.5~{\rm au}$, there is no significant mass transfer until
tidal forces bring the companions closer together on tidal timescales. To
restrict the range of periastron separations to be smaller than this limit,
the orbit cannot have a period of the order of 800 years, unless the
eccentricity is almost unity. Adopting more reasonable eccentricity values
the longest periods that still maintain a periastron distance small enough
for RLOF are of the order of a few decades.

Such a range of periods would mean that, for the detected A0V companion to
be responsible for the 800 years old nebula, not all periastron passages
result in
an interaction. This could be explained if an interaction only took place
during the last thermal pulse of the Mira (lasting $\sim$100~yr), $\sim$800
years ago. Today, the companion would be back to a smaller, inter-pulse
radius and not interacting.

However, we also found that if significant mass
transfer occurs, the orbit is altered in
such a way that within a couple more orbits the system enters a CE
interaction. The result of the CE interaction can be either a merger, or an 
AGB star that then
evolves in isolation, or a post-AGB star orbited by a
companion with a period of $\sim$1~day \citep{ivanova13}.  Since the central star of the
Rotten Egg Nebula is an AGB star, then the companion must have merged and we
are forced to conclude that
if an eccentric binary produced this nebula, then the system must have
initially been a triple:  a close
companion to the AGB star has merged with it, producing the nebula, while
the more distant A-type companion remains.  In such case the A-star would be
a by-stander. For such a triple system to be stable, the A-type
companion must be farther than $10~{\rm au}$, in line with the
observations.

In our scenario, the AGB star is the pre-interaction
primary.  Since we have no constraints on the mass of the AGB star today, we
cannot tell what the mass of the companion that merged with it was, except
that it must have been less massive than the progenitor mass of the AGB
star. Hence, the current
AGB star should be in the range $3~{\rm M_\odot}<M_{\rm AGB}<6.5~{\rm
M_\odot}$.  We do not know what consequences this merger may have had on the
AGB star, but as the merger would have occurred $\sim 800$ years ago, it is
unlikely that the AGB star would have had time to evolve away from the AGB
and it should therefore still be seen as an AGB star.

We also found that the disc-like
structure of material dragged off the AGB star during the periastron
passage of the companion that eventually merged, could be responsible for
lobes with the observed characteristics for plausible
parameters of the magnetic field structure and
by allowing for a super-Eddington
accretion rate onto the companion \citep{shiber15}.

This said we should not ignore that every phase of the evolution of a system
like ours, both before and during the CE phase is riddled with
uncertainties \citep[for a review see][]{ivanova13}. The tidal interaction
(pre-CE) phase may include enhanced, companion-induced mass-loss
\citep{tout88}, which may result in a lighter envelope at the time of CE.
The eccentricity, which should be rapidly reduced by tides, may actually be
pumped to higher values by mass loss, resulting in a system that will
interact in the way we simulate \citep{soker00}. As for whether a CE would
result after RLOF, our simulated system, with the mass
ratio of $M_{\rm d}/M_{\rm a}=1.79$ should be prone to unstable
mass transfer and CE \citep{podsiadlowski01}, although the seldom considered
case of eccentric orbits might alter the predictions. Finally, whether a
system like ours could merge inside the CE is not clear either: while there
is enough orbital energy to unbind the envelope, this criterion alone is not
sufficient to determine the outcome of the CE and depends sensitively on the
efficiency of the interaction
\citep{passy12, DeMarco2011,ivanova13}. It is possible that a lower mass
companion
may be needed to achieve a merger inside the primary's CE, but 
lower mass discs typically formed around a lower mass companion would have
more difficulty in
achieving the needed jet energies and momenta.

An eccentric binary
has been proposed to explain other objects similar to The Rotten Egg Nebula,
such as
IRAS22036+5306 \citep{sahai06,soker12}, or planetary nebulae with certain
distinctive features \citep{soker02}.  We therefore think it is worth
pursuing the current line of investigation in the context of other observed
objects, using a range of additional input parameters and with more highly
resolved simulations, which would lower some of the uncertainties.

 \section{Summary}

We have performed a number of grid-based 3D hydrodynamic simulations of
eccentric binaries consisting of an AGB star and a less massive main
sequence companion.  We found that:

\begin{itemize}
\item In order for significant mass ($\gtrsim0.01~{\rm M_\odot}$) to be 
transferred to the companion, the periastron passage must be 
$\lesssim2.5~R_{\rm star}\approx5.5~{\rm au}$ in our setup. If
significant mass transfer occurs, the orbit is altered in such a way that a
CE interaction event ensues after a few more orbits, which will
either eject the envelope and lead to a short period binary, or lead to a 
merger that retains the envelope.
\item We found that little mass is unbound in the CE
interaction, $\lesssim0.4~{\rm M_\odot}$ in all of our simulations. Most of 
the envelope therefore
remains bound.
\item Following periastron passage, but before the CE
interaction begins, an intermediate phase occurs where the companion is
surrounded by a disc-like structure that with a suitable magnetic field
can launch jets. Such a field appears likely if equipartition is assumed.
\item Provided accretion rates larger than the spherical Eddington limit and
reasonable jet launching efficiency,
the lobes in the Rotten Egg Nebula could be formed this way. However, since
that object now consists of an AGB star and a main sequence companion, and
we have concluded that the interaction leads to a merger, we suggest that if
the nebula was formed through RLOF in an eccentric binary, the
system must initially have been a triple where two objects merged to form
today's AGB star.
\end{itemize}

The next step in the study of this system and systems like it, is the use of
adaptive mesh refinement and the relaxation of the adiabatic approximation.
Simulating a triple system would incur many additional
challenges, including the need to simulate larger volumes and to run the
simulation for a longer physical time.

\section*{Acknowledgements}

We would like to thank P. Motl, M. Wardle and an anonymous referee 
for helpful discussions and comments.
J.E.S acknowledges support from the Australian Research Council Discovery 
Project (DP12013337) program.
O.D. gratefully acknowledges support from the Australian Research Council
Future Fellowship grant FT120100452.
J.-C.P. acknowledges funding from the Alexander von Humboldt Foundation. 
M.-M.M.L. was partially supported by NASA grant NNX14AJ56G and the
Alexander von Humboldt Foundation.
This research was undertaken, in part, on the National Computational
Infrastructure facility in Canberra,
Australia, which is supported by the Australian Commonwealth Government,
on the swinSTAR supercomputer at Swinburne
University of Technology and on the 
machine Kraken through grant TG-AST130034,
a part of the Extreme Science and Engineering Discovery Environment
(XSEDE), supported by NSF grant number ACI-1053575.
Computations described in this work were performed
using the \textsc{enzo} code (http://enzo- project.org), which is the product of a
collaborative effort of scientists at many universities and national
laboratories.

\appendix

\end{document}